\documentclass[twocolumn,superscriptaddress,amsfont,amssymb,amsmath, showpacs,balancelastpage, nofootinbib]{revtex4-2}
\usepackage{graphicx,longtable,mathrsfs,color,array}
\usepackage[unicode=true,pdfusetitle,
bookmarks=true,bookmarksnumbered=true,bookmarksopen=true,bookmarksopenlevel=1,
breaklinks=false,pdfborder={0 0 0},backref=false,colorlinks=true]{hyperref}
\hypersetup{citecolor=blue,filecolor=blue,linkcolor=blue,urlcolor=blue,pdfauthor={Name}
}
\usepackage[usenames,dvipsnames]{xcolor}
\usepackage{amssymb,amsmath,mathtools,mathrsfs,enumitem}
\usepackage{epsfig,subfigure,placeins,float}
\usepackage{booktabs,longtable,multirow}
\usepackage{exscale,relsize}
\usepackage[normalem]{ulem}
\usepackage[T1]{fontenc}
\usepackage[utf8]{inputenc}
\usepackage{enumerate}
\usepackage{times, mathptmx}
\usepackage{tikz}
\usepackage{aas_macros}
\usepackage{tabularx}
\usepackage{tabularray}
\usepackage{diagbox}
\usepackage{multirow}
\newcolumntype{M}[1]{>{\centering\arraybackslash}m{#1}}

\usetikzlibrary{arrows,positioning,decorations.markings,decorations.pathmorphing,calc}

\newcommand{\be}{\begin{equation}}
\newcommand{\ee}{\end{equation}}
\newcommand{\ba}{\begin{align}}
\newcommand{\ea}{\end{align}}
\newcommand{\cgw}{c_{\rm GW}}

\newcommand{\MPl}{M_{\rm Pl}}

\newcommand{\aT}{\alpha_T}

\newcommand{\comment}[1]{}

\newcolumntype{C}[1]{>{\centering\let\newline\\\arraybackslash\hspace{0pt}}m{#1}}
\newcommand*{\backin}{\rotatebox[origin=c]{-180}{$\in$}}%

\newcommand{\nocontentsline}[3]{}
\newcommand{\tocless}[2]{\bgroup\let\addcontentsline=\nocontentsline#1{#2}\egroup}

\def\Hz{{\rm Hz}}

\def\cgw{c_{\rm GW}}

\newcommand{\black}[1]{\textcolor{black}{ #1}}

\definecolor{hyperref}{RGB}{026,028,087}
\def\gsim{ \lower .75ex \hbox{$\sim$} \llap{\raise .27ex \hbox{$>$}} }
\def\lsim{ \lower .75ex \hbox{$\sim$} \llap{\raise .27ex \hbox{$<$}} }

\def\nn{\nonumber}

\newlength{\stheight}
\newcommand\textst[1][fu-grey]{
\ifmmode\setlength{\stheight}{+1.0ex}
\else\setlength{\stheight}{+0.5ex}
\fi
\bgroup\markoverwith{\textcolor{#1}{\rule[\the\stheight]{2pt}{1.0pt}}}\ULon
} 
 
\newcommand{\textins}[2][fu-grey]{
\ifmmode\mathcolor{#1}{#2}
\else\textcolor{#1}{#2}\@\,
\fi
}
\graphicspath{{./Plots/}}
\allowdisplaybreaks

\tikzstyle{vecArrow} = [thick, decoration={markings,mark=at position
1 with {\arrow[semithick]{open triangle 60}}},
double distance=1.4pt, shorten >= 5.5pt,
preaction = {decorate},
postaction = {draw,line width=1.4pt, white,shorten >= 4.5pt}]

\begin{document}

\setcounter{tocdepth}{5}
\title{Testing the Speed of Gravity with Black Hole Ringdown}

\author{Sergi Sirera}
\affiliation{Institute of Cosmology \& Gravitation, University of Portsmouth, Portsmouth, PO1 3FX, U.K.}

\author{Johannes Noller}
\affiliation{Institute of Cosmology \& Gravitation, University of Portsmouth, Portsmouth, PO1 3FX, U.K.}
\affiliation{DAMTP, University of Cambridge, Wilberforce Road, Cambridge CB3 0WA, U.K.}

\begin{abstract}
We investigate how the speed of gravitational waves, $\cgw$, can be tested by upcoming black hole ringdown observations. We do so in the context of hairy black hole solutions, where the hair is associated with a new scalar degree of freedom, forecasting that LISA and TianQin will be able to constrain deviations of $\cgw$ from the speed of light at the ${\cal O}(10^{-4})$ level from a single supermassive black hole merger. 
We discuss how these constraints depend on the nature of the scalar hair, what different aspects of the underlying physics they are sensitive to in comparison with constraints derived from gravitational wave propagation effects, which observable systems will place the most stringent bounds, and that constraints are expected to improve by up to two orders of magnitude with multiple observations.  
This is especially interesting for dark energy-related theories, where existing bounds from GW170817 need not apply at lower frequencies and where upcoming bounds from lower-frequency missions will therefore be especially powerful.
As such, we also forecast analogous bounds for the intermediate-frequency AEDGE and DECIGO missions. 
Finally, we discuss and forecast analogous black hole ringdown constraints at higher frequencies (so from LVK, the Einstein Telescope and Cosmic Explorer) and in what circumstances they can yield new information on top of existing constraints on $\cgw$.
All calculations performed in this paper are reproducible via a companion Mathematica notebook \cite{ringdown-calculations}.
\end{abstract}

\date{\today}
\maketitle

\tableofcontents

\section{Introduction} \label{sec-intro}

Measuring the speed of gravitational waves, $\cgw$, places strong constraints on the `medium' gravitational waves are propagating through and hence on the particle content of the Universe.
In the strong gravity regime, binary compact object mergers -- e.g. binary black hole (BBH) or binary neutron star (BNS) mergers -- are one of the cleanest probes of this particle content. Here interactions associated with novel particles can leave an imprint in the inspiral, merger and ringdown phases. These systems can therefore act as a particle detector, identifying or constraining the new physics that would be a consequence of such particles and associated `fifth forces'. 
One of the smoking gun signals for the presence of such new physics is a $\cgw$ different from the speed of light, and indeed it has been shown that binary compact object mergers can place powerful constraints on $\cgw$ \cite{TheLIGOScientific:2017qsa,2041-8205-848-2-L14,2041-8205-848-2-L15,LIGOScientific:2017zic,LIGOScientific:2017ync} and hence on the presence and potential dynamics of new degrees of freedom -- see e.g. \cite{Baker:2017hug,Creminelli:2017sry,Sakstein:2017xjx,Ezquiaga:2017ekz,Boran:2017rdn,LIGOScientific:2018dkp,Amendola:2012ky,Amendola:2014wma,Deffayet:2010qz,Linder:2014fna,Raveri:2014eea,Saltas:2014dha,Lombriser:2015sxa,Lombriser:2016yzn,Jimenez:2015bwa,Bettoni:2016mij,Sawicki:2016klv,Cornish:2017jml}
and references therein.
These previous constraints on $\cgw$ from binary compact object mergers have mostly focused on propagation effects (see e.g.  \cite{Baker:2017hug,Creminelli:2017sry,Sakstein:2017xjx,Ezquiaga:2017ekz} and references therein) or emission effects during the inspiral phase of such systems (see e.g. \cite{Jimenez:2015bwa}). In this paper we instead investigate what bounds can be derived on $\cgw$ from the ringdown phase alone. 
This phase is particularly amenable to being understood perturbatively and hence promises an especially clean analytic understanding.  
While strong constraints on $\cgw$ exist, most notably from the binary neutron star merger GW170817 \cite{TheLIGOScientific:2017qsa,2041-8205-848-2-L14,2041-8205-848-2-L15,LIGOScientific:2017zic,LIGOScientific:2017ync}, it is important to keep in mind that these are for frequencies in the LIGO-Virgo-KAGRA (LVK) band, i.e. $\sim 20-2000$ Hz. Expressed as an energy scale this corresponds to $\sim 10^{-14}-10^{-12}$ eV. 
This range of values is important, because dark energy-related physics is one of the primary targets that can be constrained with measurements of $\cgw$ and dark energy theories that do affect $\cgw$ generically come with a cutoff around ${\cal O}(10^2)$ Hz \cite{deRham:2018red}. This means that, for such theories, an (unknown) high energy completion of the fiducial new dark energy physics ought to take over as one approaches this cutoff, i.e. close to or somewhat below the LVK band.\footnote{Note that the cutoff is the largest possible energy/frequency scale, where the high energy completion can take over, but this can already take place at significantly lower energies/frequencies. Theoretically predicting the precise scale would require detailed knowledge about such a fiducial (currently unknown) high energy completion.} This high energy completion will naturally enforce $\cgw = c$ at high energies if it permits Lorentz invariant solutions, so LVK measurements such as GW170817 may simply confirm that feature of the high energy completion instead of probing the original (low energy) dark energy physics itself.\footnote{The same is true for bounds from (the absence of) gravitational Cherenkov radiation \cite{Moore:2001bv}, which place a lower bound on $\cgw$ at energy scales of order $\sim 10^{10}$ GeV, i.e. far above the energy scales probed by gravitational wave detectors.} 
In other words, in theories that do affect $\cgw$ at cosmological scales, one therefore naturally expects a frequency-dependent transition back to $\cgw = c$ upon approaching the LVK band.
With frequencies in the LISA band $\sim 10^{-4}-10^{-1}$ Hz (and the corresponding energies $\sim 10^{-19}-10^{-16}$ eV) being significantly lower, upcoming LISA~\cite{LISA:2017pwj} and TianQin~\cite{TianQin:2015yph} measurements therefore provide a much cleaner probe of $\cgw$ in such dark energy related theories. 
\\

{\bf Existing and upcoming constraints on $\cgw$}:
\black{Existing constraints on $\cgw$ from lower frequencies, i.e. frequencies below the LVK band, are comparatively weak, so it is particularly interesting to forecast constraints from and for LISA}. Here it is worth emphasising that a frequency-dependent $\cgw$, so e.g. different speeds in the LVK and LISA bands, is a generic consequence of the afore-mentioned dark energy theories -- see \cite{deRham:2018red,Baker:2022rhh,Harry:2022zey} for more detailed discussions on this point. 
The existing relevant constraints closest to (in fact, just below) the LISA band are from binary pulsars, in particular from the Hulse-Taylor binary, and place a bound of $|\aT| \lesssim {\cal O}(10^{-2})$ for frequencies $f \sim 10^{-5} \; \Hz$ \cite{Jimenez:2015bwa}. Here we have conveniently expressed bounds on $\cgw$ in terms of the dimensionless $\aT$ parameter
\begin{align} \label{aT_expression}
\aT \equiv (\cgw^2 - c^2)/c^2,
\end{align}
which we will use throughout this paper.
Bounds from even lower frequencies $f \sim 10^{-18} - 10^{-14} \; \Hz$ come from the cosmic microwave background and large scale structure measurements (see \cite{Noller:2018wyv,Bellini:2015xja,Hu:2013twa,Raveri:2014cka,Gleyzes:2015rua,Kreisch:2017uet,Zumalacarregui:2016pph,Alonso:2016suf,Arai:2017hxj,Frusciante:2018jzw,Reischke:2018ooh,Mancini:2018qtb,Noller:2018eht,Arai:2019zul,Brando:2019xbv,Arjona:2019rfn,Raveri:2019mxg,Perenon:2019dpc,Frusciante:2019xia,Arai:2019zul,SpurioMancini:2019rxy,Bonilla:2019mbm,Baker:2020apq,Joudaki:2020shz,Noller:2020lav,Noller:2020afd,Traykova:2021hbr}
and references therein) and require $|\aT| \lesssim {\cal O}(1)$. Finally there are already a number of $\cgw$-related forecasts for upcoming measurements in the LISA band: 
\begin{itemize}
    \item[1)] \cite{Littenberg:2019mob} forecasted that a multi-messenger observation 
    in the LISA band using observations of an eclipsing white-dwarf binary will be able to constrain $|\aT| \lesssim 10^{-12}$ (in the event of a non-detection of any $\aT$-related effect). 
    \item[2)] For the case when there is a significant frequency-dependence for $\cgw$ already within the LISA band, \cite{Baker:2022rhh} used redshift-induced frequency dependence imprinted on waveforms to be observed in the LISA band (i.e. without the need for an optical counterpart) to forecast a constraint of $|\aT| \lesssim 10^{-4}$. 
    \item[3)] Also for frequency-dependent $\cgw$ within the LISA band, \cite{Harry:2022zey} forecasted that a bound $|\aT| \lesssim 10^{-17}$ can be placed by using the fact that waveforms to be observed by LISA will be squeezed/stretched/scrambled due to the different speeds with which different frequencies will propagate (for frequency-dependent $\cgw$ and again without the need for an optical counterpart).
    \item[4)] Finally, if there is no detectable frequency-dependence in both the LISA or LVK bands individually, but a transition in between, \cite{Harry:2022zey,Baker:2022eiz} showed that multiband observations using systems such as GW150914 -- that are first observable in the LISA band and later enter the LVK band \cite{Sesana:2016ljz} -- will constrain $|\aT| \lesssim 10^{-15}$ (again in the event of a non-detection).\footnote{\black{Several of the forecasted constraints listed here were computed considering single waveforms. For upcoming future detectors signal overlap will likely be a regular occurrence, so understanding to what extent this impacts the above bounds will be an interesting issue to explore going forward.}}    
\end{itemize}
Looking forward to upcoming LISA observations this leaves us with the following situation when looking for the strongest possible upcoming bounds. If there is any significant frequency-dependence in the LISA band, a strong $|\aT| \lesssim 10^{-17}$ bound 
will very quickly be established once a single sufficiently loud SMBH (super massive black hole) merger has been observed. No optical counterpart or multi-band observation is required for this. If no such frequency-dependence is present, multi-messenger events 
and multi-band observations will eventually place bounds at the $10^{-12}$ level and $10^{-15}$ level, respectively.\footnote{Note that the galactic eclipsing white dwarf binary considered in \cite{Littenberg:2019mob} is a known system which is expected to be clearly observable in LISA, whereas detection rates for multi-messenger events more akin to GW170817 (i.e. compact object mergers with a clearly identifiable optical counterpart that pinpoints the merger itself) are highly uncertain \cite{Amaro-Seoane:2022rxf}. Multi-band observations as discussed in \cite{Sesana:2016ljz,Harry:2022zey,Baker:2022eiz} will take several years to constrain $\cgw$, given the signal has to `migrate' from the LISA to the LVK frequency band for the constraint to arise.}  
Here we will show that additional bounds at the $10^{-4}$ level can be derived from the ringdown phase of an observed SMBH merger. These bounds are more model-dependent (we will detail how below), but will effectively be obtainable as soon as LISA goes online (given an expected ${\cal O}(10-100)$ observable SMBH mergers per year 
\cite{Berti:2006ew,Sesana:2004gf,Rhook:2005pt,Tanaka:2008bv,Berti:2009kk,eLISA:2013xep,Bonetti:2018tpf}
). 
In the LISA context these bounds are therefore most relevant in the event that no significant frequency-dependence is detectable within the LISA band itself, e.g. when $\cgw$ quickly asymptotes to a constant value for high and low frequencies and its frequency-dependence and transition between those asymptotes is effectively localised to a narrow band between the frequencies accessible by LISA and LVK. We will further discuss this setup -- as pointed out above, this is the same basic setup as explored in \cite{Harry:2022zey,Baker:2022eiz} in the context of multi-band observations -- below, as well as how our analysis is affected when frequency-dependence leaks into the frequency band under investigation.
As we will show, the ringdown bounds on $\cgw$ discussed here are also eventually expected to tighten by up to two orders of magnitude when stacking observations of multiple events. We will also highlight that such bounds are not just a complementary and independent constraint on $\cgw$, but the fact that they are derived for a different background space-time compared with constraints from gravitational wave propagation (black hole vs. cosmological space-times) also allows us to extract novel insights about the underlying physics.
%
\\

{\bf Scalar-tensor theories}: We will focus on theories where the fiducial new physics is minimal in the sense that it is described by a single scalar degree of freedom $\phi$, so that we are dealing with a scalar-tensor theory. The most general such theory which results in second-order equations of motion is commonly know as Horndeski scalar-tensor theory \cite{Horndeski:1974wa,Deffayet:2011gz}\footnote{For the equivalence between the formulations of \cite{Horndeski:1974wa} and \cite{Deffayet:2011gz}, see \cite{Kobayashi:2011nu}.}, which is governed by the following action
\begin{align}
    S= \int &d^4x \sqrt{-g}\Big[G_2 + G_3 \Box \phi 
    + G_4 R + \nn \\
    &G_{4X}\left[(\square\phi)^2-\phi^{\mu\nu}\phi_{\mu\nu}\right] +G_5G_{\mu\nu}\phi^{\mu\nu}- \nn \\ &\frac{1}{6}G_{5X}\left[(\square\phi)^3 -3\phi^{\mu\nu}\phi_{\mu\nu}\square\phi+2\phi_{\mu\nu}\phi^{\mu\sigma}\phi_\sigma^\nu\right]\Big].
    \label{S}
\end{align}
Here we have introduced the shorthands $\phi_\mu\equiv\nabla_\mu\phi$ and $\phi_{\mu\nu}\equiv\nabla_\nu\nabla_\mu\phi$, and the $G_i$ are free functions of $\phi$ and $X$, where $X\equiv-\frac{1}{2}\phi_\mu\phi^\mu$. $G_{iX}$ denotes the partial derivative of $G_i$ with respect to $X$. 
Most relevant for our purposes will be the ($X$-dependent parts of the) $G_4$ interactions and the $G_5$ interactions, since (as we will discuss below) these are the only interactions affecting $\aT$. 
Also note that, for simplicity, we will often focus on the case where $G_4$ is $\phi$-independent, so that $G_{4\phi} = 0$ (where $G_{i\phi}$ denotes the partial derivative of $G_i$ with respect to $\phi$) -- we discuss what this assumption entails in more detail in appendix \ref{app-full-param}. 

It is important to highlight that, just like General Relativity (GR), the Horndeski scalar-tensor theory \eqref{S} is an effective field theory (EFT), so has a limited range of validity. 
When \eqref{S} is taken to be a fiducial dark energy theory that does affect $\cgw$ on cosmological scales, then this theory only applies up to its cutoff, expected at or below the aforementioned ${\cal O}(10^2)$ Hz.   
This is precisely analogous to the way in which GR is at most a valid description of gravitational phenomena up to the Planck scale. 
These observations have an important practical implication when computing BBH merger observables as we do here: 
$\cgw$ and hence $\aT$ derived from \eqref{S} are in fact frequency-independent as a consequence of the structure of \eqref{S} imposed by the requirement of 2nd order equations of motion. The frequency-dependence of $\cgw$ alluded to above only enters as a consequence of the unknown UV (high energy) completion of \eqref{S}, in other words once we are about to leave the regime of validity of \eqref{S}. Throughout most of this paper we will compute and analyse ringdown predictions derived from \eqref{S}, so we are implicitly assuming that we are operating within a frequency-window where I) \eqref{S} is firmly within its regime of validity, and hence II) $\cgw$ is effectively frequency-independent {\it within} this window. 
Rigorously computing analogous predictions in frequency-windows where there is significant frequency-dependence for $\cgw$ would require incorporating at least some of the effects of the UV completion and hence supplementing/replacing \eqref{S} with the relevant interactions.
We will point out the implications of this assumption in more detail below as well as when one can extrapolate to more general scenarios.  
\\

{\bf Outline}: With the above setup in place, this paper is organised as follows. We collect and discuss the relevant results from black hole perturbation theory in section \ref{sec:BHpert}, both for `bald' and `hairy' black hole solutions. We extract the observable quasinormal spectrum from the relevant solutions in section \ref{sec:QNMs}, discussing issues related to the parametrisation of $\aT$ in the process. Parametrized constraints are then presented in section \ref{sec:param-constraints}, where we analytically compute the precision with which upcoming ringdown observations will be able to probe $\aT$ for a generic observation. We discuss correlations between different constrainable parameters and how constraints depend on the underlying interactions. Forecasted constraints for a range of specific missions and experiments are then discussed in section \ref{sec:forecast-constraints}. We conclude in section \ref{sec:conclusions} and collect further relevant details in the appendices.

\section{Black hole perturbation theory} \label{sec:BHpert}
Since the ringdown phase of BBH mergers can be well-described perturbatively, we first ought to discuss the relevant setup in black hole perturbation theory. 
We will consider static and spherically symmetric background solutions that are Ricci-flat ($R_{\mu\nu}=0=R$) here, in particular Schwarzschild spacetimes. We therefore write the background metric $\bar g_{\mu\nu}$ as
\begin{equation}
    ds^2=\bar g_{\mu\nu}dx^{\mu}dx^{\nu}=-A(r)dt^2+\frac{1}{B(r)}dr^2+C(r)d\Omega^2,
\end{equation}
where $A,B,C$ are general functions of the radial coordinate $r$ and $d\Omega^2$ is the line element of the standard 2-sphere. We now consider metric perturbations $h$ around this background, where
\begin{align}
    g_{\mu\nu}&=\bar g_{\mu\nu}+h_{\mu\nu}.
\end{align}
Around the static and spherically symmetric backgrounds considered here such perturbations can be decomposed into odd and even parity perturbations (under rotations), which decouple from one another at linear order \black{(i.e. they evolve independently from one another and can therefore be treated separately)}.
We will work to leading (linear) order in this paper, but note that this decoupling does not hold at higher orders -- see \cite{Brizuela:2006ne,Brizuela:2007zza,Brizuela:2009qd,Lagos:2022otp,Mitman:2022qdl,Cheung:2022rbm} for details on the behaviour of higher order modes.
In this paper we will exclusively focus on odd perturbations, which can be written as
\begin{equation}
    h_{\mu\nu}^{\rm odd}=\left(\begin{array}{cccc}
        0 & 0 & 0 & h_0\\
        0 & 0 & 0 & h_1\\
        0 & 0 & 0 & 0\\            
        h_0 & h_1 & 0 & 0
    \end{array} \right)\sin{\theta}\partial_\theta Y_{\ell m}
    \label{hodd}
\end{equation}
where we have used the Regge-Wheeler gauge \cite{ReggeWheeler} and, since we assume a static background metric, we have set $m=0$ without loss of generality. $h_0$ and $h_1$ are functions of $(r,t)$, where the $t$-dependence will be taken to be of the form $e^{-i\omega t}$.
Since perturbations of the scalar $\phi$ are even under parity transformations and we focus on parity odd modes, we will therefore only be concerned with metric perturbations. These perturbations are however affected by the background solution they are propagating on, so odd metric perturbations will nevertheless be sensitive to the new physics encoded by the (background solution of the) fiducial scalar degree of freedom we are probing here.

\subsection{Schwarzschild black holes without hair}

No-hair theorems guaranteeing a trivial scalar field profile exist for a wide range of scalar-tensor theories \cite{hawking1972black,PhysRevLett.108.081103,PhysRevLett.110.241104,Sotiriou:2013qea}.\footnote{This was first shown for stationary black holes in minimally coupled Brans-Dicke theories \cite{hawking1972black}, and subsequently extended to a more general class of scalar-tensor theories including self-interactions of the scalar \cite{PhysRevLett.108.081103}, to spherically symmetric static black holes in Galilean-invariant theories \cite{PhysRevLett.110.241104}, and for slowly rotating black holes in more general shift-symmetric theories \cite{Sotiriou:2013qea}.} A natural starting point are therefore Schwarzschild spacetimes with a constant scalar field background profile $\bar\phi$ as e.g. investigated by \cite{Tattersall:QNMsHorn,Tattersall:BHspectro,Motohashi:2018wdq}
\begin{align}
ds^2&=-\left(1-\tfrac{2M}{r}\right)dt^2+\frac{1}{\left(1-\tfrac{2M}{r}\right)}dr^2+d\Omega^2, \nn \\
\bar\phi & = {\rm constant}.
\label{backg}
\end{align}
Note that, when we mention the `background' or `background solution' going forward, we refer to {\it both} the metric {\it and} scalar background solutions, as e.g. provided in \eqref{backg}.
Around the background \eqref{backg} odd metric perturbations trivially behave just as in GR, since they are unaffected by the even sector (where scalar perturbations do induce non-trivial effects) and also do not feel any effects from the scalar background solution (since this is trivial in the present no-hair setup).
So in order to explore potentially observable effects induced by the scalar, one ought to either investigate different background solutions or consider even perturbations. For detailed discussions of the second option we refer to \cite{Kobayashi:2014wsa, Tattersall:QNMsHorn,Tattersall:BHspectro,Tattersall:GenBHPert,Franciolini:2018uyq,Datta:2019npq,Khoury:2020aya,Bernardo:2020ehy,Langlois:2022eta,Minamitsuji:2022mlv,Hui:2021cpm} for work in the context of Horndeski gravity, and to
\cite{Tattersall:GenBHPert,deRham:2019gha,Langlois:2021aji,Glampedakis:2019dqh,Chen:2021pxd,Chen:2021cts} for work in the context of other theories (scalar-tensor or otherwise). 
However, here we will proceed along the first route, considering the dynamics of odd perturbations around different background solutions.
\black{We leave an investigation of how the speed of gravity impacts quasi-normal modes in the even sector in the presence of a non-trivial background solution (i.e. combining the two options discussed above) for future work}.

\subsection{Hairy black holes: Background}

If $\phi$ acquires a non-trivial background profile, this will provide a medium for gravitational waves (i.e. here in particular $h^{\rm odd}_{\mu\nu}$) to travel through and hence can affect $\cgw$. Probing $\cgw$ therefore constitutes a powerful test for departures from GR in such cases, as neatly illustrated in the aforementioned cosmological context. For the black hole solutions we focus on here, a well-known scalar-tensor theory example that can have scalar hair are scalar-Gauss-Bonnet (sGB) theories \cite{Sotiriou:2013qea,Sotiriou:2014pfa}.
In the context of Horndeski theories these are described by an action that (in addition to a standard kinetic terms) contains a $G_5$ interaction where $G_5 \sim \ln |X|$ \cite{Kobayashi:2011nu}. However, instead of focusing on a specific hairy solution, we will here follow the approach of \cite{Tattersall:QNMHair} and parametrise the scalar-induced hair in a perturbative fashion, but otherwise remain agnostic about the precise nature of the hair.  
More specifically, we will consider a no-hair Schwarzschild black hole solution at lowest order and introduce small hairy deviations away from this. These can in principle manifest themselves both in the background solution for the metric as well as in the scalar profile, so \cite{Tattersall:QNMHair} proposed the following parametrised ansatz
\begin{align}
A(r) &= B(r) = 1-\frac{2M}{r} + \epsilon \delta A_1 + \epsilon^2 \delta A_2 + {\cal O}(\epsilon^3), \nn \\    
C(r) &= (1 + \epsilon \delta C_1 + \epsilon^2 \delta C_2)r^2 + {\cal O}(\epsilon^3) \nn \\
\bar\phi &= \widehat\phi + \epsilon \delta \phi_1 + \epsilon^2 \delta \phi_2 + {\cal O}(\epsilon^3).
\label{hair_parametrisation}
\end{align}
Here $\delta A_i, \delta C_i, \delta \phi_i$ are functions of $r$ and $\epsilon$ is simply a useful order parameter, since we will work perturbatively up to quadratic order in the (small) hair $\delta\phi$ -- $\epsilon$ has no physical meaning beyond this.\footnote{It is worth emphasising an important subtlety here. As mentioned above, since we are focusing on the odd parity sector of perturbations, there are no scalar perturbations contributing in our setup. The $\delta\phi_i$ in \eqref{hair_parametrisation} therefore describe small deviations in the background solution for the scalar away from a lowest order constant scalar profile. There are therefore implicitly two perturbative hierarchies at play here. Metric perturbations $h_{\mu\nu}^{\rm odd}$ and perturbations in the background hair. We will work up to linear order in $h_{\mu\nu}$ (at the level of the equations of motion) and up to quadratic order in the hair.} Note that we will denote quantities which are evaluated on the background (so $h_{\mu\nu}$ is set to zero, recall odd scalar perturbations vanish identically) with a bar, so $\bar\phi$ denotes the scalar field as evaluated on the background. Quantities where in addition the small (background) scalar hair $\delta\phi_i$ is set to zero are denoted by a hat, so e.g. $\widehat\phi$ denotes the scalar field as evaluated on the background in the absence of any non-trivial scalar hair. As a consequence we have $\widehat X =0$, while $\bar X$ here acquires non-zero contributions via the $\delta\phi_i$. 
While \eqref{hair_parametrisation} is a very general parametrisation, for our purposes we will be able to work with a highly simplified subset. We are interested in probing the effect of $\cgw$ (or equivalently $\aT$) on the ringdown phase. Since deviations from $\cgw = c$ (or equivalently $\aT = 0$) arise due to a non-trivial scalar field profile acting as a medium for gravitational waves passing through, it is unsurprising that at lowest order in $\epsilon$ any $\aT$-dependent contribution only depends on $\delta\phi_1$ in \eqref{hair_parametrisation} and not on $\delta A_i, \delta C_i$, or $\delta \phi_2$. We collect results showing this explicitly in appendix \ref{app-full-param}, but here we will therefore proceed by working with the much simpler parametrised ansatz
\begin{align}
ds^2&=-\left(1-\tfrac{2M}{r}\right)dt^2+\frac{1}{\left(1-\frac{2M}{r}\right)}dr^2+d\Omega^2, \nn \\
\bar\phi &= \widehat\phi + \epsilon \delta \phi.
\label{hair_parametrisation_simple}
\end{align}
Recall that $\delta\phi$ is a small deviation in the background solution $\bar\phi$. This will allow us to identify the leading order contributions imprinted by a non-zero $\aT$, so is ideally suited for our purposes. We will later discuss to what extent the constraints we will derive on $\aT$ may be contaminated/weakened in the presence of non-zero $\delta A_i, \delta C_i$, but for now proceed with \eqref{hair_parametrisation_simple} as a proof of principle. However, do note that our simplified ansatz is (partially) motivated by sGB-like hair. There, when working perturbatively in a small sGB coupling, at leading order only the scalar background acquires a non-trivial contribution, while the metric remains Schwarzschild \cite{Sotiriou:2013qea,Sotiriou:2014pfa},\footnote{This motivation is only `partial', since 1) it would correspond to setting $\delta A_1 = 0 = \delta C_1$ in \eqref{hair_parametrisation}, but not $\delta A_2, \delta C_2$ and 2) because the guiding principle here is not to explore the consequences of any specific theory, but rather to explore the consequences of a non-trivial $\cgw$ on top of a parametrised background ansatz. See appendix \ref{app-full-param} for a more in-depth discussion of what happens when $\delta A_1 = 0 = \delta C_1$, but $\delta A_2, \delta C_2$ are non-zero and fully taken into account.} i.e. we are working with a so-called `stealth' solution for the metric.

\subsection{Hairy black holes: Quadratic action}

In order to extract the ringdown signal we need to compute the behaviour of (odd parity) perturbations on top of the background \eqref{hair_parametrisation_simple}. Working out the quadratically perturbed action, substituting the components of \eqref{hodd} as well as our background solution \eqref{hair_parametrisation_simple}, integrating over the angular coordinates and performing several integrations by parts, we recover the action \cite{Kobayashi:2012kh,Ganguly:2017ort}
\begin{align}
    S^{(2)}=\int dtdr\left[\bar a_1h_0^2+\bar a_2h_1^2+\bar a_3\left(\dot{h}_1^2+h_0'^2-2\dot{h}_1h_0'+\frac{4}{r}\dot{h}_1h_0\right)\right],
    \label{S2v2}
\end{align}
where a dot and a prime denote derivatives with respect to $t$ and $r$, respectively, and we have dropped an overall multiplicative factor of  $2\pi/(2\ell +1)$ coming from  angular integration. The expressions for the $\bar a_i$ agree with those found by \cite{Kobayashi:2012kh,Ganguly:2017ort} and satisfy
\begin{align}
    &\bar a_1=\frac{\ell(\ell+1)}{2r^2}\left[\left(r\mathcal{H}\right)'+\frac{(\ell-1)(\ell+2)\mathcal{F}}{2B}+\frac{r^2}{B}\epsilon_A\right],\nn\\
    &\bar a_2=-\frac{\ell(\ell+1)}{2}B\left[\frac{(\ell-1)(\ell+2)\mathcal{G}}{2r^2}+\epsilon_B\right],\nn\\
    &\bar a_3=\frac{\ell(\ell+1)}{4}\mathcal{H}.
\end{align}
where the $\bar a_i$ are to be evaluated on the background \eqref{hair_parametrisation_simple} (to avoid clutter bars are implied, but not written explicitly, for all expressions on the right hand side). $\epsilon_{A,B}$ are contributions that vanish on-shell, and
\begin{align}
    \mathcal{F}&=2\left(G_4+\frac{1}{2}B\phi'X'G_{5X}-XG_{5\phi}\right),\nn\\
    \mathcal{G}&=2\left[G_4-2XG_{4X}+X\left(\frac{B'}{2}\phi'G_{5X}+G_{5\phi}\right)\right],\nn\\
    \mathcal{H}&=2\left[G_4-2XG_{4X}+X\left(\frac{B}{r}\phi'G_{5X}+G_{5\phi}\right)\right].
    \label{FGH}
\end{align}
The quadratic action (\ref{S2v2}) contains two fields $(h_0,h_1)$, but describes only one dynamical degree of freedom. \cite{Kobayashi:2012kh,Ganguly:2017ort} show how the action can be rewritten to make this manifest. To do so an auxiliary field $q$ is defined and then re-defined into a field $Q$, satisfying\footnote{In short, the one-field quadratic action is obtained by introducing an auxiliary field $q$ into \eqref{S2v2} while leaving the dynamics invariant, varying with respect to $h_{0,1}$ to obtain $h_{0,1}$ in terms of $q$, substituting them back into the action and finally performing a field redefinition $q(Q)$. This is explicitly shown in the accompanying notebook \cite{ringdown-calculations}.}
\begin{align}
    h_0 &= -\frac{(r^2\bar a_3q)'}{r^2 \bar a_1-2(r \bar a_3)'}, &h_1 &=\frac{\bar a_3}{\bar a_2}\dot{q}, &q &= \frac{\sqrt{\mathcal{F}}}{r\mathcal{H}}Q.
\end{align}
Re-writing the quadratic action in terms of $Q$ in tortoise coordinates $r_*$ (defined as $dr=Bdr_*$), one then finds
\begin{align}
    S^{(2)}=\frac{\ell(\ell+1)}{4(\ell-1)(\ell+2)}\int dtdr_*\left[\frac{\mathcal{F}}{\mathcal{G}}\dot{Q}^2-\left(\frac{dQ}{dr_*}\right)^2-V(r)Q^2\right],
    \label{S2Q}
\end{align}
where the potential is given by 
\begin{align}
    V &= (\ell+2)(\ell-1)\frac{B}{r^2}\frac{\mathcal{F}}{\mathcal{H}}-\frac{r^3}{2}\left(\frac{B^2}{r^4}\right)'-\frac{r^4\mathcal{F}^2}{4\mathcal{F}'}\left(\frac{B^2\mathcal{F}'^2}{r^4\mathcal{F}^3}\right)'.
    \label{V}
\end{align}
Note that, we have liberally used our background ansatz \eqref{hair_parametrisation_simple} to simplify the $\bar a_i$, etc. in comparison to the more general expressions in \cite{Kobayashi:2012kh,Ganguly:2017ort} -- we collect those general expressions in appendix \ref{app-acoefficients} for comparison.

\subsection{Modified Regge-Wheeler equation}
In order to obtain the analogue of the Regge-Wheeler equation, we now vary the action with respect to the field $Q$ and find
\begin{equation}
    \frac{\partial^2Q}{\partial r_*^2}-\frac{\mathcal{F}}{\mathcal{G}}\frac{\partial^2Q}{\partial t^2}-VQ=0.
\end{equation}
We assume that the time dependence of Q is given by $e^{-i\omega t}$, substitute $\mathcal{F}$, $\mathcal{G}$, $\mathcal{H}$ and $V$ for our background \eqref{hair_parametrisation_simple}, and finally obtain the modified Regge-Wheeler equation
\begin{equation}
    \frac{d^2Q}{dr_*^2}+\Biggl[\omega^2(1+\epsilon^2\alpha_T)-B(V_{RW}+\epsilon^2\delta V)\Biggr]Q=0,
    \label{Rweq}
\end{equation}
where $\alpha_T$ satisfies
\begin{equation}
\alpha_T=-B\frac{G_{4X}-G_{5\phi}}{G_4}\delta\phi'^2.
\label{aTexp}
\end{equation}
Note that, given the background we are considering, $\alpha_T$ naturally is a function of $r$ as well as of the Schwarzschild mass $M$. In general there are further contributions to $\aT$ depending on $G_{5X}$,
but these only enter at cubic order in $\epsilon$ as can be deduced from \eqref{FGH}, so do not contribute here.\footnote{This also implies that \eqref{aTexp} can be simplified further by integrating the contributing $G_{5}$ terms by parts in the original theory, but as we will work with a general $\aT$ here, this does not affect our subsequent expressions.} $V_{RW}$ is the well-known Regge-Wheeler potential in GR
\begin{equation}
    V_{RW}=\frac{\ell(\ell+1)}{r^2}-\frac{6M}{r^3},
\end{equation}
and $\delta V$ is given by
\begin{equation}
    \begin{split}
     \delta V=\alpha_T&\Biggl[\frac{M(2r-5M)}{r^3(r-2M)}+\frac{(\ell+2)(\ell-1)}{r^2} \\
     &-\frac{r-2M}{2r}\left(\left(\frac{\delta\phi''}{\delta\phi'}\right)^2-\frac{\delta\phi'''}{\delta\phi'}\right)+\frac{r-5M}{r^2}\frac{\delta\phi''}{\delta\phi'}\Biggr].  
    \end{split}
    \label{dV}
\end{equation}
While it has been re-arranged into a more concise form here, this as well as the above expressions in this subsection agree with the corresponding results given in \cite{Tattersall:QNMHair}, when specialised to our ansatz \eqref{hair_parametrisation_simple}.
Note that we have implicitly assumed that $G_{4\phi}=0=G_{4\phi\phi}$ here, as would e.g. be the case in shift-symmetric theories. We do this to isolate the effect of $\aT$ on the ringdown spectrum,
but will further discuss how $G_{4\phi}\neq0\neq G_{4\phi\phi}$ would affect our results in appendix \ref{app-full-param}.

\section{Quasinormal modes}\label{sec:QNMs}

Having derived and collected the relevant results from black hole perturbation theory in the previous section, we are now in a position to extract the key observable in the black hole ringdown context: the quasinormal modes (QNMs). 
As before, we will be focusing on the perturbations of odd modes and the modified Regge-Wheeler equation \eqref{Rweq} governing them. This equation can now be solved to obtain the frequencies of the associated quasinormal modes $\omega$. Unlike normal modes, these frequencies are complex numbers, where the real part represents the physical oscillation frequency and the imaginary part represents the exponential damping due to dissipation in the system.
The QNM spectrum only depends on the properties of the final black hole (mass, angular momentum, charge) as well as on the structure of the underlying theory. Detecting and measuring this spectrum is hence a powerful way to constrain the presence of novel degrees of freedom and interactions, as well as to generally test the Kerr hypothesis \cite{Berti:2016lat},\footnote{The hypothesis states that the spacetime around a black hole after gravitational collapse is well described by the Kerr metric and therefore contains no hair.} a scheme that has received the name of black hole spectroscopy. Note that, while the QNM spectrum does not depend on initial conditions, the amplitude of individual modes does and this will be relevant for us in section \ref{sec:param-constraints}.
\\

\subsection{Parametrized ringdown}\label{subsec-ParametrizedRingdown}
There are a number of techniques one can use to obtain the QNM themselves (see e.g. \cite{Ferrari:1984zz,1972ApJ...172L..95G,1985ApJ...291L..33S,Motl:2003cd,Motl:2003cd,Horowitz:1999jd,Berti:2009wx,Leaver:1985ax}) but we refer to \cite{Berti:2009kk} for an extensive review of those.\footnote{The rationale for obtaining numerical solutions for QNM is schematically the following. One imposes boundary conditions on the horizon and spatial infinity (corresponding to $r_*\rightarrow\pm\infty$) such that on the horizon wavepackets are moving inwards and at infinity wavepackets are moving outwards. The imposition of boundary conditions will then select `quantised' values of $\omega$ (poles in the Green's function) which correspond to the QNM.} In this paper, we will make use of the parametrized ringdown formalism \cite{ParamRingdown}, the relevant key aspects of which we will now summarise.
In order to apply this formalism, our modified Regge-Wheeler equation has to be re-cast into the following form
\begin{equation}
    B\frac{d}{dr}\left(B\frac{dQ}{dr}\right)+\Bigl[\omega^2-B(V_{RW}+\tilde{\delta V)}\Bigr]Q=0.
    \label{RWeqParam}
\end{equation}
This can easily be achieved by absorbing the $\epsilon^2\omega^2\alpha_T$ term into $\delta V$ in (\ref{Rweq}). Because this term is a small correction to $\tilde{\delta V}$, we can take $\omega$ to be the unperturbed frequencies $\omega_0$ of the unmodified Regge-Wheeler equation characteristic of GR, around which we will compute the leading order $\delta\omega$ corrections below.\footnote{As we will clarify more explicitly in the next section, there are multiple modes encoded within $\omega_0$.} Doing so, we obtain a modified Regge-Wheeler equation in the form of (\ref{RWeqParam}) with
\begin{equation}
    \tilde{\delta V}=\epsilon^2\left(\delta V-\frac{1}{B}\omega_0^2\alpha_T\right),
    \label{dVtilde}
\end{equation}
with $\delta V$ being given by (\ref{dV}). It is then instructive to express the modification to the potential $\tilde{\delta V}$ as an expansion in powers of $(2M/r)$
\begin{equation}
    \tilde{\delta V}=\frac{1}{(2M)^2}\sum_{j=0}^{\infty}a_j\left(\frac{2M}{r}\right)^j.
    \label{dV1}
\end{equation}
Once expressed in this form, \cite{ParamRingdown} show that the quasinormal frequencies are determined by the same $a_j$ coefficients as follows
\begin{equation}
    \omega=\omega_0+\delta\omega \equiv \omega_0 + \sum_{j=0}^\infty a_j e_j,
    \label{domega}
\end{equation}
given that a smallness criterion on the coefficients $|a_j|\ll(1+1/j)^j(j+1)$ is satisfied. The $e_j$ are a complex `basis' and we summarise the low-order $e_j$ most relevant here in table \ref{e-basis} -- for more details and an explicit computation of this basis see \cite{ParamRingdown}.

At this point we can already appreciate an important subtlety from the structure of equations \eqref{dV1} and \eqref{domega}. 
Each coefficient $a_j$ contributing to $\tilde{\delta V}$ enters with different (increasing) powers of $\sim 1/r$ (\ref{dV1}). 
While this does mean that those contributions to the potential are suppressed in the far distance limit, i.e. far away from the horizon, it does not entail that these contributions are providing a sub-dominant contribution to the frequency spectrum for the QNMs.
Indeed, from \eqref{domega} we explicitly see that this $\sim 1/r^j$ suppression does not play a role in determining the QNM frequencies. The $j$-th correction enters as $a_je_j$ and, while the $e_j$'s tend to slowly decrease in size as $j$ increases, there is no parametric suppression of higher $j$ contributions. 
Also note that situations where (some) higher $j$ contributions dominate over lower $j$ contributions do arise rather generically -- we will see explicit examples below. Finally, note that the smallness criterion mentioned above guarantees that the $j$-th contribution to the QNM frequencies is a small correction to $\omega_0$, but this does not entail that the sum of all corrections has to be parametrically suppressed.

\subsection{Parametrizing scalar hair and \texorpdfstring{$\aT$}{[aT]}} \label{subsec-ParametrizedHair}
\begin{figure}[t!]
    \centering
    \includegraphics[width = 0.48\textwidth]{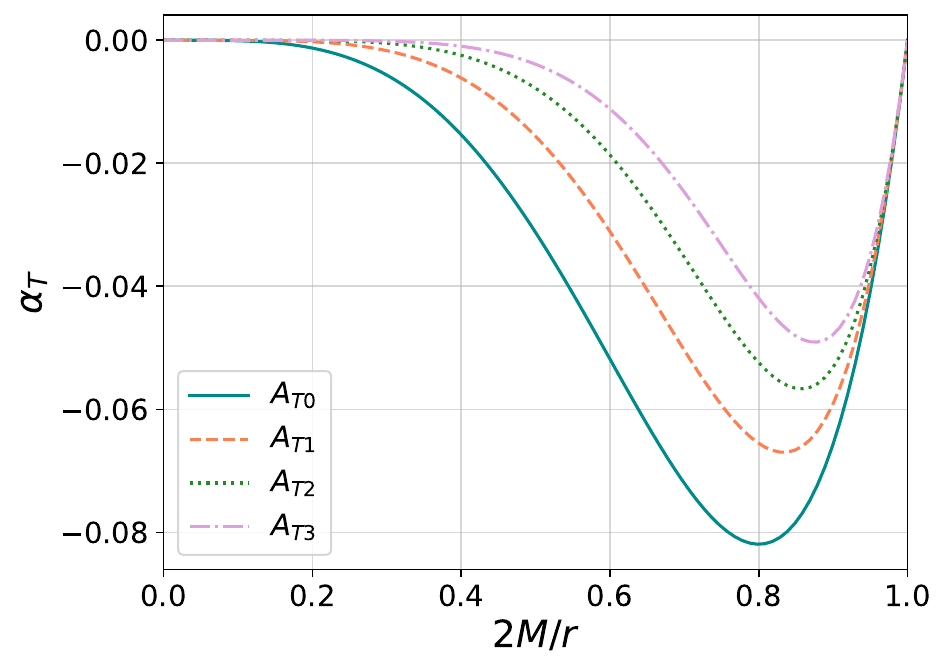}\\[0.5cm]
    \caption{Here we plot $\alpha_T$, the deviation of the speed of gravitational waves from that of light defined via $\aT \equiv (\cgw^2 - c^2)/c^2$, as a function of $2M/r$. 
    We show $\aT$ for different example choices of $i$ in \eqref{aTfull}, where only the amplitude $A_{Ti}$ corresponding to this specific $i$ is non-zero (and fixed to a fiducial value  of $1$) for each plotted curve.
    Note that $2M/r=0$ therefore corresponds to spatial infinity whereas $2M/r=1$ corresponds to the Schwarzschild radius, so we are interested in this range of values/distances. 
    One can clearly see that $\aT(r) = 0$ at spatial infinity and at the horizon, but displays non-trivial behaviour in-between with the overall amplitude and sign determined by $A_{Ti}$ while the radial dependence depends on the choice of $i$ considered. 
    }
    \label{fig-aT}
\end{figure}
Before proceeding with the QNM computation and applying the above formalism to our \eqref{dVtilde}, we require more information about the functional form of $\delta\phi$ and $\aT$. In close analogy to the above discussion, it is natural to to view these functions as an expansion in powers of $(2M/r)$ as well.
Starting with the scalar hair function $\delta\phi$, in the main text we will follow \cite{Tattersall:QNMHair} and focus on a scalar hair profile parametrised as
\begin{equation}
    \delta\phi=\varphi_c\left(\frac{2M}{r}\right),
    \label{dphi}
\end{equation}
where $\varphi_c$ is a constant.\footnote{Note that such a profile is indeed recovered in scalar Gauss-Bonnet theories \cite{Sotiriou:2013qea,Sotiriou:2014pfa} (in the long distance limit and when working perturbatively in the Gauss-Bonnet coupling).}
In appendix \ref{app-generaln} we discuss the more general parametrisation $\delta\phi=\varphi_c\left(\frac{2M}{r}\right)^n$ (where one remains agnostic of the leading order in $1/r$ at which the hair enters) as well as superpositions of different $r$-dependencies in the scalar hair profile.
We leave an investigation of even more general (non-power-law) parametrisations for future investigation.
Note that, while in this section we will focus on the $n=1$ scalar hair profile \eqref{dphi}, we will discuss how different profiles affect eventual constraints on $\aT$ in section \ref{subsec-HairProfile}.
Having parametrised $\delta\phi$, we turn our attention to the one remaining function of $r$ affecting $\tilde{\delta V}$, namely $\aT$.
To this end it will be useful to separate out the dependence on the scalar hair background profile and other geometric factors from $\alpha_T$ in \eqref{aTexp} as follows
\begin{align}
    \alpha_T &= -B(2M)^2G_{T}\delta\phi'^2,
    &G_{T}&\equiv \frac{1}{(2M)^2}\frac{G_{4X}-G_{5\phi}}{G_4}.
    \label{alT}
\end{align}
Here the dimensionless $G_T$ parameter has been defined to isolate the dependence of $\aT$ on the Lagrangian $G_i$ functions, as opposed to the $r$-dependence following directly from the scalar profile or via the dependence on the Schwarzschild function $B(r)$. We will find this separation especially useful later on when investigating what constraints on $\aT$ can tell us about scalar hair and vice versa.\footnote{Note that we have used the Schwarzschild mass $M$ as a mass scale to define a dimensionless  $G_T$ here, but in principle this mass scale is arbitrary.}
Because we have a non-trivial scalar profile, all the $G_i$ (and hence also  $G_T$) are functions of $r$ and so to fully specify the $r$-dependence of $\aT$ we finally also expand  $G_T$ in powers of $r$ along the same lines as discussed for $\delta\phi$ above. Doing so we can write
\begin{equation}
     G_T=\sum_iG_{Ti}\left(\frac{2M}{r}\right)^i,
    \label{alTi}
\end{equation}
where each of the $G_{Ti}$ are constant coefficients. Putting everything together, i.e. substituting $\delta\phi$ \eqref{dphi} and $G_{T}$ \eqref{alTi} into the expression for  $\alpha_T$ \eqref{alT}, we can finally write 
\begin{align}
    \alpha_T&=-\sum_{i=0}^\infty G_{Ti}\varphi_c^2\left(1-\frac{2M}{r}\right)\left(\frac{2M}{r}\right)^{i+4}, \nn \\
    &= -\sum_{i=0}^\infty A_{Ti}\left(1-\frac{2M}{r}\right)\left(\frac{2M}{r}\right)^{i+4}.
    \label{aTfull}
\end{align}
In the final line we have implicitly defined a final shorthand as part of our notational setup. The dimensionless amplitude parameters $A_{Ti}$ neatly encapsulate the coefficients controlling $\aT$ and satisfy
\begin{align}
    A_{Ti} \equiv G_{Ti}\varphi_c^2.
    \label{ATi_def}
\end{align}
As one may expect, these are also the effective constant parameters that, as we will find below, QNM observations will constrain observationally.
To provide some intuition on the relationship between $A_{Ti}$ and $\aT$, 
we illustrate the dependence of $\aT$ on $2M/r$ for various choices of the amplitude coefficients $A_{Ti}$ in figure \ref{fig-aT}.

\subsection{Parametrized QNMs} \label{subsec-ParametrizedQNMs}
Having parametrized all the functional freedom encoded within $\tilde{\delta V}$ above, it is now straightforward to combine the above expressions. Doing so we can express $\tilde{\delta V}$ as
\begin{align}
   \tilde{\delta V}=\frac{1}{(2M)^2}\sum_{i=0}^\infty A_{Ti}\Bigg[&\left(\frac{2M}{r}\right)^{4+i}(2M\omega_0)^2 \nn\\
    +&\left(\frac{2M}{r}\right)^{6+i}\left(-\ell(\ell+1)+9\right)\nn\\
    +&\left(\frac{2M}{r}\right)^{7+i}\left(\ell(\ell+1)-20\right)\nn\\
    +&\left(\frac{2M}{r}\right)^{8+i}\frac{45}{4}\Bigg],
    \label{dVn1}
\end{align}
Note that we have dropped the order parameter $\epsilon^2$ at this point. From this we can read off the $a$-coefficients defined in (\ref{dV1})\footnote{For a given $i$ the contributions to these coefficients are
\begin{align}
    &a_{4+i}\ \ \backin\ \ A_{Ti}(2M\omega_0)^2,\nn\\
    &a_{6+i}\ \ \backin\ \ A_{Ti}(-\ell(\ell+1)+9),\nn\\
    &a_{7+i}\ \ \backin\ \ A_{Ti}(\ell(\ell+1)-20),\nn\\
    &a_{8+i}\ \ \backin\ \ A_{Ti}\frac{45}{4}.
    \label{aij}
\end{align}
By using the `element sign' $\in$ we stress that a given $a_j$ can be built from contributions from different $i$'s. For instance, $a_6$ obtains contributions from $i=0$ and $i=2$.}
and from \eqref{domega} we also obtain the following expression for the quasinormal frequencies
\begin{align}
    \delta\omega&=\sum_{i=0}^\infty A_{Ti}\cdot E_i^1,\nn\\
    &=\sum_{i=0}^\infty A_{Ti}\Big[(2M\omega_0)^2e_{4+i}-(\ell(\ell+1)-9)e_{6+i}\nn \\
    &\hspace{1.5cm} +(\ell(\ell+1)-20)e_{7+i}+\frac{45}{4}e_{8+i}\Big],
    \label{domega_v2}
\end{align}
where $E_i^1$ has been defined for convenience, and its subscript $1$ refers to $n=1$. A set of $E_i^n$ `basis' functions for general $n$ 
are provided in appendix \ref{app-generaln}.

Anticipating some of our later discussion, we can already see that, for small $\ell$ and $2M\omega_0 \sim {\cal O}(1)$, the higher $e_j$ terms are enhanced relative to smaller $j$ terms (for a given $A_{Ti}$). For additional details on how different contributions enter into $\delta\omega$, see appendix \ref{app-generaln} and especially figure \ref{triangle}. There we also explicitly discuss how this picture changes for different scalar field profiles. Also note that the consistency criterion we alluded to above, $|a_j|\ll(1+1/j)^j(j+1)$, places an implicit bound on the amplitudes $A_{Ti}$ for the perturbative treatment we have outlined to be valid. As an example, for the case where only $i=0$ terms contribute this bound requires that $A_{T0} \ll 1.5$.\footnote{Note that in practice this bound is set by the $a_7$ coefficient, which places a stronger bound than the other $a_j$.}
Including higher order $i$'s will generate similar joint constraints on different $A_{Ti}$.
As we will see, observational bounds will constrain the $A_{Ti}$ at the $10^{-1}$ level or stronger. Given we are measuring deviations away from the standard GR expectation $A_{Ti} = 0$, we therefore expect these consistency bounds to be satisfied in all relevant scenarios here.
\\
\begin{table}[!t]
 \setlength{\tabcolsep}{10pt}
     \centering
     \begin{tabular}{ccc}\toprule
                        & Re($2Me_j$) & Im($2Me_j$)\\\midrule
          $j=4$         & 0.03668 & -0.00044\\
          $j=5$         & 0.02404 & 0.00273\\ 
          $j=6$         & 0.01634 & 0.00484\\
          $j=7$         & 0.01136 & 0.00601\\
          $j=8$         & 0.00795 & 0.00654\\\bottomrule
     \end{tabular}
     \caption{Real and imaginary components of the $e_j$ `basis' functions for $\ell=2$, taken from \cite{ParamRingdown}. Note that we start with $j=4$ as this is our lowest-order non-zero $a_j$. For the full collection up to $j=50$ for each $\ell$ up to $\ell=10$, together with the `basis' for even-gravitational and even-scalar perturbations, see \cite{ParamRingdown}.}
     \label{e-basis}
 \end{table}

\section{Parametrized constraints}\label{sec:param-constraints}

In the previous section we derived an analytic expression for the QNM frequencies, assuming these to be close to the corresponding GR frequencies for a Schwarzschild black hole with the small perturbations encoding information about interactions in the underlying scalar-tensor theory, in particular about $\aT$.
We would now like to use this to forecast how well future GW experiments will be able to constrain $\aT$ using ringdown.
More specifically, we perform a Fisher forecast to estimate the error in the $A_{Ti}$ parameters \eqref{ATi_def}. In this section we therefore derive general expressions for the resulting constraints and discuss their overall features, following this up in the next section by forecasting and discussing constraints for specific upcoming experiments. 
Note that, throughout this section, we ubiquitously use the techniques developed in \cite{Berti_2006} for our analysis.

\subsection{Fisher forecast setup}

\begin{table}[!t]
 \setlength{\tabcolsep}{10pt}
     \centering
     \begin{tabular}{ccc}\toprule
                        & Re($2M\omega_0$) & Im($2M\omega_0$)\\\midrule
          $\ell=2$         & 0.7474 & -0.1779\\
          $\ell=3$         & 1.1989 & -0.1854\\ 
          $\ell=4$         & 1.6184 & -0.1883\\
          $\ell=5$         & 2.0246 & -0.1897\\\bottomrule
     \end{tabular}
     \caption{Real and imaginary components of the quasinormal frequencies $\omega_0$ of a Schwarzschild BH in GR for $\ell=2,3,4,5$. Quasinormal data is provided online \cite{grit,berti-ringdown}.}
     \label{SchwQNM}
 \end{table}
We begin by modelling the waveform as
\begin{equation}
    h=h_+F_++h_\times F_\times,
\end{equation}
where $h_{+,\times}$ represent the strain in the two polarisations of the gravitational wave. These are in principle functions of all coordinates, i.e. $h_{+,\times}(t,r,\theta,\phi)$. However, to distinguish between time and frequency domains, we will only make $t$ (or $\nu$) explicit and take the dependence on $r,\theta,\phi$ as understood. $F_{+,\times}$ are functions encoding the geometry of the problem (i.e. they depend on the angles specifying the orientation of the source with respect to the detector).
The strain functions for the ringdown are given by
\begin{align}
    h_+(t)&=\sum_{\ell m}A_{\ell m}^+e^{-\frac{\pi t f_{\ell m}}{Q_{\ell m}}}S_{\ell m}\cos{(\phi^+_{\ell m}+2\pi t f_{\ell m})},\nn\\
    h_\times(t)&=\sum_{\ell m}A_{\ell m}^+N_\times e^{-\frac{\pi t f_{\ell m}}{Q_{\ell m}}}S_{\ell m}\sin{(\phi^\times_{\ell m}+2\pi t f_{\ell m})},
    \label{strain}
\end{align}
where these are the strain functions as emitted by the source and we will implicitly assume that these  trivially propagate to the detector here and will briefly discuss what this assumption entails and when propagation effects can be relevant in the next section.
In \eqref{strain} we have absorbed any overall constant normalization factors into the amplitude parameters $A_{\ell m}^{+}$,\footnote{The strain functions $h_{+/\times}$ appear with different normalisation factors in the literature depending on the setup in question, e.g. with a factor of $1/2\sqrt{10\pi}$, an extra geometrical $\sqrt{3/4}$ for LISA or a $\frac{1}{r}$ factor \cite{Berti_2006,London:2014cma,Berti:2007zu,Berti:2016lat}. We choose to remain general and absorb all such factors into the amplitudes $A^+$. This does not affect the calculations presented in this section, as these factors only enter trough the signal-to-noise-ratio $\rho$, for which the appropriate detector-specific values will be discussed and used in the next section.
} and where $f_{\ell m}$ and $\tau_{\ell m}$ characterise the real and imaginary parts of $\omega_{\ell m}$ in the following way
\begin{align}
    &\omega_{\ell m}=2\pi f_{\ell m}+\frac{i}{\tau_{\ell m}}, & &Q_{\ell m}=\pi f_{\ell m}\tau_{\ell m},
    \label{ftQ}
\end{align}
where $Q_{\ell m}$ is the quality factor. $\{A^+_{\ell m},A^\times_{\ell m}=A^+_{\ell m}N_\times,\phi^+_{\ell m},\phi^\times_{\ell m}\}$ are the amplitudes and phases for the two polarisations.  Finally, $S_{\ell m}$ are spheroidal functions carrying angular dependencies.
Because modes with different $(\ell,m)$ do not mix due to the nature of our background, the $\ell m$ indices in $\{\omega_{\ell m},f_{\ell m},\tau_{\ell m},Q_{\ell m},S_{\ell m},A^+_{\ell m},\phi^+_{\ell m},\phi^\times_{\ell m}\}$ will play no role for the time being, so we will obviate them to simplify our notation (and explicitly discuss which $\ell m$ modes are of interest when this becomes relevant below).

Using the above strain functions, we compute the signal-to-noise-ratio (SNR) with the usual 
\begin{equation}
    \rho^2=(h|h)=4\int_0^\infty d\nu\frac{\tilde{h}(\nu)^*\tilde{h}(\nu)}{S_h(\nu)},
\label{rho2}
\end{equation}
where $S_h(\nu)$ is the noise spectral density characteristic of the detector and $\tilde{h}(\nu)$ is the Fourier transform of $h(t)$.\footnote{Note that in \eqref{rho2} we use $\nu$ rather than $f$ for the frequency domain representation (or Fourier transform) of the time coordinate. This is to distinguish it from the real component of the quasinormal modes $f_{\ell m}$ as defined in \eqref{ftQ}, especially since we will be omitting the $\ell m$ indices.}
We now make use of the following set of simplifying assumptions: $\langle F_+\rangle=\langle F_\times\rangle=1/5$, $\langle F_+F_\times\rangle=0$, $\langle|S|^2\rangle=1/4\pi$, $A^+=A$. We also make use of the fact that we can approximate $S_h(\nu)$ to be constant. For details on (and explicit checks of) these assumptions see \cite{Berti_2006}.
Using these assumptions, \eqref{rho2} can be re-expressed as\footnote{Note that to obtain this simple expression for $\rho^2$ we have further assumed that $\phi^+=\phi^\times$ and $N_\times=1$ (i.e. $A^\times=A^+=A$) \cite{Berti_2006}. However, we stress that this is not a necessary assumption to recover the expression for the single-parameter error (\ref{error}) laid out in the following subsections. We do find it necessary to recover the double-parameter error expressions (\ref{errors}). Without this assumption, we have
\begin{equation}
    \rho^2=\frac{QA^2((1+N_\times^2)(1+4Q^2)+\cos{2\phi^+}-N_\times^2\cos{2\phi^\times})}{(1+N_\times^2)(1+4Q^2)\pi fS_h}.
\end{equation}
}
\begin{equation}
    \rho^2=\frac{QA^2}{\pi fS_h}.
    \label{SNR}
\end{equation}
To derive error estimates we make use of the Fisher Information Matrix, given by
\begin{equation}
    \Gamma_{ab}=\left(\frac{\delta h}{\delta\theta^a}\Big|\frac{\delta h}{\delta\theta^b}\right),
    \label{FM}
\end{equation}
where $\theta^a$ is the set of parameters for our theory and the noise-weighted product $(\cdot|\cdot)$ is defined as
\begin{equation}
    (h_1|h_2)=2\int_0^\infty d\nu\frac{\tilde{h_1}^*\tilde{h_2}+\tilde{h_2}^*\tilde{h_1}}{S_h(\nu)}.
\end{equation}
Then, we can calculate the parameter errors by inverting the Fisher matrix (which gives the covariance matrix $\Sigma$). 
The error for a parameter $a$ is given by
\begin{equation}
    \sigma_a=\sqrt{\Sigma_{aa}}=\sqrt{\Gamma_{aa}^{-1}}.
\end{equation}

As an initial estimate we will here study the simplified case where all the usual parameters of the waveform are known $(A,\phi^+,...)$ and our only free parameters are the $A_{Ti}$'s \eqref{ATi_def}. \black{We leave forecasting full joint constraints to future work. This simplified setup means the estimates which we will compute below effectively are upper bounds on the precision one can expect.}
\comment{We leave forecasting full joint constraints to future work, but point out that this simplified setup means the estimates which we will compute below effectively are upper bounds on the precision one can expect.}
For a setup as considered here, where the only waveform parameters we want to constrain are those appearing inside the quasinormal frequencies $\omega$ (i.e. inside $f$ and $Q$)\footnote{Note that we need two variables to represent the real and imaginary parts of $\omega$ and, similarly to previous literature \cite{Tattersall:QNMHair} we choose to work with the pair $\{f,Q\}$ rather than $\{f,\tau\}$. We note that, if working with the latter, the SNR equation (\ref{SNR}) would be further simplified to
\begin{equation}
    \rho^2=\frac{\tau A^2}{S_h}.
\end{equation}
}, \black{general expressions for the errors can be analytically derived. These only depend on the number of parameters one wants to constrain.} In this paper we will constrain up to two $A_{Ti}$ together so we provide here the expression for single-parameter constraints\footnote{This simple expression also implicitly makes use of the large-$Q$ limit (or equivalently large damping times $\tau$). More terms appear at the $Q^{-4}$ order. For details on the validity of this approximation we again refer to \cite{Berti_2006} and the full `unapproximated' expressions are available in the companion notebook \cite{ringdown-calculations}.}
\begin{equation}
    \sigma_{A_{Ti}}^2\rho^2=\frac{1}{2}\left(\frac{f}{Qf'}\right)^2,
    \label{error}
\end{equation}
where the prime denotes a derivative with respect to $A_{Ti}$, and for double-parameter constraints
\begin{align}
    &\sigma_{A_{Ti}}^2\rho^2=\frac{\dot{f}^2}{2}\frac{(2Q)^2+(1-\frac{f\dot{Q}}{\dot{f}Q})^2}{(\dot{Q}f'-\dot{f}Q')^2},\nn\\
    &\sigma_{A_{Tj}}^2\rho^2=\frac{f'^2}{2}\frac{(2Q)^2+(1-\frac{f Q'}{f'Q})^2}{(\dot{Q}f'-\dot{f}Q')^2},
    \label{errors}
\end{align}
where again a prime denotes a derivative with respect to $A_{Ti}$, and a dot represents a derivative with respect to $A_{Tj}$. This matches analogous expressions in \cite{Tattersall:QNMHair} (albeit for non-$\aT$-related parameters there).

Before deriving error estimates on different parameter combinations, let us briefly return to the question of which $(\ell,m)$ modes are of most interest.\footnote{There is a third index characterising the QNM spectrum, the overtone number $n$. Here we only focus on the `fundamental mode' $n=0$. Modes with higher $n$'s (i.e. overtones) are more suppressed by virtue of having increasing values of $|\text{Im}(\omega)|$.} As discussed above, while the QNM spectrum does not depend on initial conditions, the amplitude of individual modes does. 
The dominant observable contributions, i.e. the modes with the largest amplitudes for astrophysical binary compact object mergers, generically are the
$\ell=m$ modes, more specifically the $(2,2)$ mode \cite{Berti_2006,London:2014cma,Berti:2007fi,Berti:2007zu,Bhagwat:2019bwv,Bhagwat:2019dtm}.\footnote{Note that the discussion in section \ref{sec:BHpert} only applies for $\ell \geq 1$ modes and the dipole perturbation $\ell=1$ requires special treatment, as the Regge-Wheeler gauge used in section \ref{sec:BHpert} does not fully fix all gauge degrees of freedom for this mode, see \cite{Kobayashi:2012kh} for details. However, the contribution from the $\ell=1$ mode can be shown to be negligible for the background solutions we are probing \cite{Kobayashi:2012kh}, so this is of no concern here.}
Note that, for a non-rotating black hole solution as we are focusing on here, the equations of motion are independent of $m$ \cite{Tattersall:GenBHPert}. So while $m=0$ is typically fixed in such setups for simplicity, as we have done here, the results derived apply for any $m$.
The relative amplitude of subdominant modes (in particular $\ell=3$) grows as the mass ratio $q$ and angular momentum $j$ of the remnant black hole increases \cite{Berti:2007zu,London:2014cma,Kamaretsos:2011um,Baibhav:2020tma} -- \black{also see those references for discussions related to the detectability of such modes}. Nonetheless, the $\ell=2$ mode still generically dominates in all scenarios and higher $\ell$ modes decay more quickly, see table \ref{SchwQNM}. Note that the damping time $\tau$ goes as the inverse of the imaginary component of $\omega$, which increases for higher $\ell$ modes. So in addition to generically possessing a smaller amplitude, these modes also decay faster.
Finally, also notice that, for binary systems that have orbited each other for a sufficiently long time for orbits to have approximately circularised, the $\ell = 2$ mode will be additionally enhanced relative to other modes \cite{LIGOScientific:2018jsj,PhysRev.136.B1224,PhysRevD.77.081502}. While it is straightforward to repeat the analysis for other higher $\ell$ modes, \black{the above rationale truly singles out the $\ell=2$ mode as the observationally most relevant. We will therefore focus on this mode in what follows.}
Having said this, a multiple mode analysis will of course be a powerful tool to probe higher dimensional parameter spaces using ringdown alone tests in the future.
While, as we have seen, quasinormal modes are independent of $m$ for static black holes, all astrophysical black holes do in fact rotate. For those, $(2,2)$ is truly the dominant mode, and hence we will focus on this one to perform the Fisher forecast and leave a more detailed study of constraints in the presence of several detected modes for the future. Extending the quasinormal mode calculations in sections \ref{sec:BHpert} and \ref{sec:QNMs} to rotating black holes is an interesting way forward for which some machinery already exists, at least for slowly-rotating black holes (see e.g. \cite{Tattersall:KerrdSBH,Pani:2012bp,doi:10.1142/S0217751X13400186,Pani:2013wsa,Brito:2013wya}). However, such metrics and the Schwarzschild metric are smoothly connected (i.e. taking the limit of zero rotation $j\rightarrow0$ recovers the Schwarzschild line element) so one expects that the non-rotating scenario still captures the leading order information in the quasinormal frequencies for sufficiently slow rotation.

\subsection{Constraining \texorpdfstring{$A_{T0}$}{[aT0]}}
We begin by considering a minimal setup, where there is only a single relevant $A_{Ti}$ parameter, namely $A_{T0}$. From \eqref{domega_v2} we then find the QNM shift to be given by
\begin{align}
    \delta\omega=A_{T0}\cdot E_0^1.
    \label{domega_AT0}
\end{align}
where $E_0^1$ is shown in \eqref{domega_v2} and we quote it here for reference
\begin{align}
    E_0^1=\Big[&(2M\omega_0)^2e_{4}-(\ell(\ell+1)-9)e_{6}\nn \\
    &+(\ell(\ell+1)-20)e_{7}+\frac{45}{4}e_{8}\Big].
\end{align}
Substituting in the numerical values for the $e_j$ from table \ref{e-basis}, we obtain
\begin{equation}
    M\delta\omega=-[0.00070+0.00306i]\cdot A_{T0}.
    \label{Mdome}
\end{equation}
In evaluating this, we have also used the $\ell=2$ mode in table \ref{SchwQNM}. This now allows us to obtain parametric expressions for the $\alpha_T$-induced deviations in the QNM spectrum. From \eqref{Mdome} and table \ref{SchwQNM} we find the following percentage differences for the real and imaginary parts, respectively
\begin{align}
    &\frac{\delta\omega_R}{\omega_{0R}}\approx-0.19\cdot A_{T0}\%, & &\frac{\delta\omega_I}{\omega_{0I}}\approx3.44\cdot A_{T0}\%.
\end{align}
Finally, we are also in a position to extract an expression for the accuracy with which an experiment with ringdown SNR $\rho$ will be able to measure $A_{T0}$. Reading off $f$ and $Q$ from \eqref{Mdome}, as defined in \eqref{ftQ}, and substituting them into the single-parameter error expression \eqref{error}, we obtain an estimate on its detectability in the same fashion as \cite{Tattersall:QNMHair}.\footnote{Note that, in evaluating the final expression, we set $A_{T0}$ to zero. This should simply be understood as capturing the leading order contributions to the error -- depending on the actual value of $A_{T0}$ the precise error can differ by $ \lesssim {\cal O}$(10\%).}
This gives us\footnote{More precise results are provided in \cite{ringdown-calculations}. Ultimately, we will only be interested in the robust order-of-magnitude constraints here, so e.g. in table \ref{tab-SNRringdown} we will approximate $\sigma_{A_{T0}}\rho\approx {\cal O}(10^2)$.}
\begin{equation}
    \sigma_{A_{T0}}\rho\approx 181.
\end{equation}
\black{The numerical value of this error calculation, as well as the analogous ones which will follow in this section, will be translated into $\aT$ constraints for specific detectors in section \ref{sec:forecast-constraints}. There, we will compare our results with other existing and forecasted constraints.}
\subsection{Constraining multiple \texorpdfstring{$A_{Ti}$}{[aTi]}}\label{multATi}
Having considered the single-parameter case above, a natural next step is to consider a more complex functional form for $\aT$ and hence for the $A_{Ti}$. Here we consider the case where $\aT$ is controlled by two parameters, $A_{T0}$ as before and a second parameter $A_{T1}$. Proceeding as before, we then have
\begin{align}
    \delta\omega=A_{T0}\cdot E_0^1+A_{T1}\cdot E_1^1.
    \label{dome01}
\end{align}
Reading off expressions for $f$ and $Q$ 
from equation \eqref{dome01} as before, we find the following error estimates from \eqref{errors} 
\begin{align}
    \sigma_{A_{T0}}\rho &\approx 302,
    &\sigma_{A_{T1}}\rho &\approx 465.
\label{ai2-con}
\end{align}
One may wonder how we can constrain two parameters with the measurement of a single mode. To this end note that the measurement of a single mode carries information about the oscillation frequency of that mode as well as for the associated damping time and these independent pieces of information allow constraining two parameters here. Once future observations are capable of measuring multiple modes \cite{Bhagwat:2019bwv,Berti:2009kk,Baibhav:2019rsa}, this will of course allow constraining a correspondingly larger parameter space. 

Equation \eqref{ai2-con} shows that $A_{T0}$ and $A_{T1}$ can be constrained to a similar order of precision. This re-iterates that terms at higher order in a $1/r$ expansion are not parametrically suppressed in their contribution to the QNM frequency spectrum, so a $1/r$ expansion is not an ideal basis in terms of observational constraints.
Indeed, upon closer inspection, we find that constraints on $A_{T0}$ and $A_{T1}$ are strongly correlated, as can be seen from the off-diagonal elements of the covariance matrix for these two parameters
\begin{equation}
    \Sigma_{ab} \sim 10^4\begin{pmatrix}
9 & -16\\
-16 & 22
\end{pmatrix},
\label{CM}
\end{equation}
We can therefore diagonalize the covariance matrix to obtain the eigenmodes that will be constrained by the data, i.e. a more optimal basis from a detectability point of view.\footnote{We thank Sigurd Naess for related discussions.} Under standard matrix diagonalization procedures we obtain
\begin{equation}
    \tilde{\Sigma}_{ab}=S^{-1}\Sigma_{ab}S \sim 10^3\begin{pmatrix}
5 & 0\\
0 & 303
\end{pmatrix}.
\label{CMAB}
\end{equation}
This transformation amounts to identifying the combinations of $A_{T0}$ and $A_{T1}$ that yield uncorrelated parameters $A_{TA}$ and $A_{TB}$, i.e. we have performed the parameter transformation $(A_{T0},A_{T1})\rightarrow(A_{TA},A_{TB})$ such that the covariance matrix of the latter is the one given by equation (\ref{CMAB}). More explicitly, the relevant eigenmodes here are $A_{TA} = -0.84 A_{T0} - 0.54 A_{T1}$ and $A_{TB} = -0.54 A_{T0} + 0.84 A_{T1}$.\footnote{Equivalently, the matrix $S$ is built with the eigenvectors of (\ref{CM})
\begin{equation}
    S=\begin{pmatrix}
\vec{s_1} & \vec{s_2}
\end{pmatrix} \sim \begin{pmatrix}
-0.84 & -0.54\\
-0.54 & 0.84
\end{pmatrix}.
\end{equation}
}
Finally, the errors for the new parameters are
\begin{align}
    \sigma_{A_{TA}}\rho &\approx 68,
    &\sigma_{A_{TB}}\rho &\approx 550,
\end{align}
where we indeed see that $A_{TA}$ can be constrained more tightly than any parameter in the previous basis.

\subsection{Dependence on scalar hair profile}
\label{subsec-HairProfile}
\begin{figure}[t!]
    \centering
    \includegraphics[width = 0.48\textwidth]{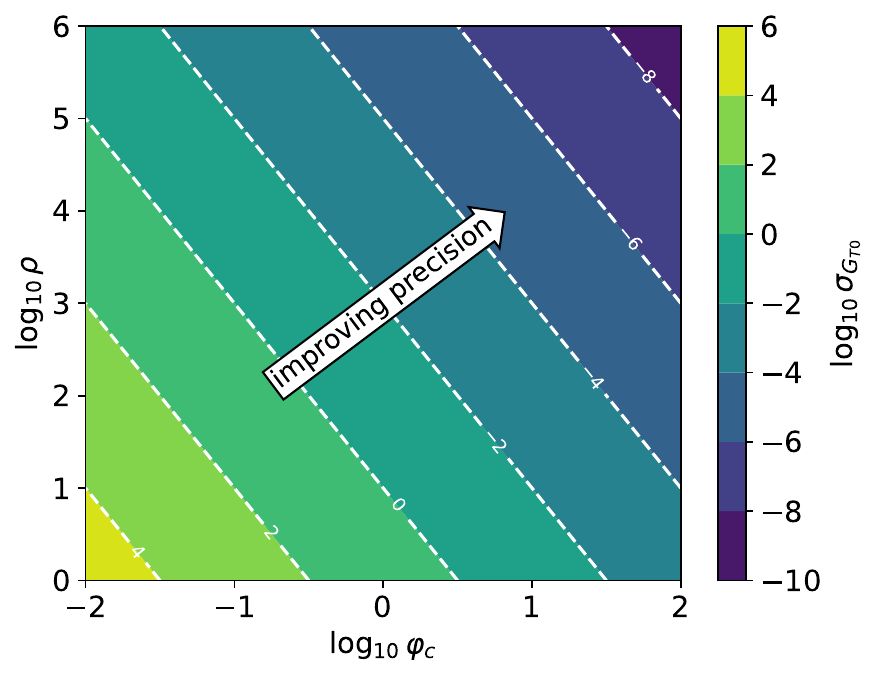}\\[0.5cm]
    \caption{Here we show forecasted errors on the strength of the interactions contributing to $\cgw$ as parametrised by $G_{Ti}$ \eqref{alTi}, where we focus on $G_{T0}$ as an example. The corresponding error $\sigma_{G_{T0}}$ is shown as a function of the scalar hair amplitude $\varphi_c$ and of the detector SNR as quantified by $\rho$. We see that $\sigma_{G_{T0}}$ improves as the scalar hair amplitude grows and as $\rho$ increases, as expected from \eqref{sigmaGT}.
    More concretely, for an SNR of $\rho \sim 10^x$ and a scalar hair amplitude of order $\varphi_c \sim 10^y$, we find $\sigma_{G_{T0}} \sim 10^{2-x-2y}$. From table \ref{tab-SNRringdown}, at lower frequencies for LISA we would therefore expect $\sigma_{G_{T0}} \sim 10^{-3-2y}$ constraints, while for LVK one would need $y \gtrsim 8$ to yield constraints on the underlying interactions there that are competitive with or stronger than existing bounds in this band.}
    \label{errorSNRphi}
\end{figure}
Above we have derived expressions for the precision with which a generic future experiment with SNR $\rho$ will be able to constrain the relevant parameter combinations affecting QNM frequencies, namely the $A_{Ti}$. Here we would like to investigate to what extent the specific form of the scalar hair profile affects this. As we will argue, in certain cases this argument can then also be inverted to place constraints on the scalar hair itself. Recall that we are parametrising the scalar hair profile as  
\begin{equation}
    \delta\phi=\varphi_c\left(\frac{2M}{r}\right)^n.
    \label{dphi_gen}
\end{equation}
Here $\varphi_c$ effectively captures the scalar field amplitude, while $n$ carries information about the radial dependence of this profile. Until now we have set $n=1$.
\\

{\bf Amplitude}:
The QNM frequencies derived above are functions of the $A_{Ti}$, which we recall depend both on the scalar amplitude $\varphi_c$ as well as on the $G_{Ti}$ (i.e. the interactions in the underlying theory) via \eqref{ATi_def}. This has an immediate important consequence, namely that a detection of the specific QNM shifts discussed here implies {\it both} a detection of scalar hair and of non-trivial $G_4$ and/or $G_5$ interactions contributing to the $G_{Ti}$ -- cf. \eqref{alT}. 
The scalar amplitude $\varphi_c$, analogously to the amplitudes of QNMs, will depend on the `initial conditions' for the ringdown phase. It is worth emphasising that, at present, it is not yet well understood how the non-linear merger stage affects this amplitude in scalar-tensor theories of interest, so we will leave $\varphi_c$ as a free parameter.
\footnote{$\varphi_c$ may be significantly enhanced or suppressed during the non-linear merger stage, so in the absence of comprehensive numerical (merger) simulations for the theories in question, even an order of magnitude estimate appears premature. Note that, in cases where the scalar hair does affect the black hole geometry (so unlike the `stealth' solutions \eqref{hair_parametrisation_simple} we consider here), this effect on the geometry can be used to place additional constraints on the nature and amplitude of the hair e.g. along the lines presented in \cite{Khodadi:2020jij,Antoniou:2022dre,EventHorizonTelescope:2020qrl,EventHorizonTelescope:2021dqv,Vagnozzi:2022moj}.}
It is interesting, then, to disentangle the effect of $\varphi_c$ and of $G_{Ti}$ on the constrained $A_{Ti}$ parameter(s).
This is shown in Figure \ref{errorSNRphi}. As can clearly be seen, and indeed as expected from  \eqref{ATi_def}, in the presence of a larger scalar hair amplitude $\varphi_c$ the constraint on the $G_{Ti}$ becomes stronger. More explicitly
\begin{align}
    &\sigma_{G_{Ti}} = \sigma_{A_{Ti}}\varphi_c^{-2}
    \label{sigmaGT}
\end{align}
Interestingly, this implies that one can I) infer a constraint on the scalar hair amplitude from measurements of the QNMs, {\it given} another non-trivial bound on the $G_{Ti}$ e.g. from non-ringdown related constraints on $\cgw$ (in other words: in the event of a future detection of a $\cgw \neq c$ from another probe), and II) infer a constraint of the $G_{Ti}$, {\it given} other independent information about the amplitude $\varphi_c$ (in other words: in the event of a complementary detection of scalar hair). 
\\

{\bf Radial dependence}:
Having considered the effect of the scalar hair amplitude above, we now investigate how our analysis is affected when the functional form of the scalar hair, i.e. its $r$-dependence and hence $n$ in \eqref{dphi_gen}, changes. 
We have considered $n=1$ above and here we repeat the Fisher analysis for $n=2$ as a complementary example -- for further details on the effect of general $n$ see appendix \ref{app-generaln}. For concreteness, we again consider the single-parameter $A_{T0}$ case. The corrections to the quasinormal frequencies are now given by
\begin{align}
    \delta\omega=A_{T0}\cdot E_0^2,
\end{align}
where for $n=2$ the basis $E_i^n$ \eqref{E-basis} becomes
\begin{align}
    E_i^2=4\Bigg[&(2M\omega_0)^2e_{6+i}+\left(-\ell(\ell+1)+\frac{31}{2}\right)e_{8+i}\nn\\
    &+\left(\ell(\ell+1)-\frac{69}{2}\right)e_{9+i}+\frac{76}{4}e_{10+i}\Bigg],
    \label{En2}
\end{align}
leading to the following expression for the error on $A_{T0}$
\begin{equation}
    \sigma_{A_{T0}}\rho \approx 132.
\end{equation}
We see that the precision with which $A_{T0}$ can be measured has improved for the $n=2$ case compared to $n=1$. We find this tightening of constraints with increasing $n$ to be generic and discuss it more in appendix \ref{app-generaln}, also for the two parameter case with $A_{T0}$ and $A_{T1}$.

\section{Forecasting observational constraints}\label{sec:forecast-constraints}
\begin{table}[!t]
\centering
\begin{tblr}{
  colspec = {|c|c|c|}
}
\hline[1pt]
   Detector(s) & Ringdown SNR ($\rho$) & Error on $\aT$\\
   \hline[1pt]\hline[1pt]
    LVK & $10$ \cite{TheLIGOScientific:2016src,Flanagan:1997sx,Nakano:2021bbw} 
    & $1$\\
   \hline[1pt]
       ET / CE & $10^2$ \cite{Maggiore:2019uih,Nakano:2021bbw,Evans:2021gyd,Hall:2022dik} & $10^{-1}$\\
   \hline[1pt]\hline[1pt]
    pre-DECIGO & $10^2$ \cite{Nakamura:2016hna} & $10^{-1}$\\
   \hline[1pt]
    DECIGO / AEDGE & $10^3$ \cite{Nair:2018bxj,AEDGE:2019nxb}* & $10^{-2}$\\
    \hline[1pt] \hline[1pt]
    LISA & $10^5$ \cite{Flanagan:1997sx,Zhang:2021kkh} & $10^{-4}$\\
   \hline[1pt]
    TianQin & $10^5$ \cite{Zhang:2021kkh} & $10^{-4}$\\
   \hline[1pt]
    AMIGO & $10^5$ \cite{Baibhav:2019rsa} & $10^{-4}$\\
   \hline[1pt]
\end{tblr}
\caption[ringdownSNR]{Achievable order-of-magnitude ringdown SNRs for a single observed event for different GW detectors and the corresponding order-of-magnitude errors on $\aT$.
Errors in this table are computed assuming $A_{T0}$ is the only amplitude parameter contributing to $\aT$ in \eqref{aTfull}, as an example. The error on $\aT$, $\sigma_{\aT}$, is quoted as one order-of-magnitude less than the corresponding error on $A_{T0}$, as observed in figure \ref{fig-aT}. We stress that the precise mapping of underlying amplitude parameters to $\aT$ mildly depends on the precise functional form of the scalar hair and the underlying interactions, but note that errors on other $A_{Ti}$ parameters and hence $\aT$ are qualitatively similar -- see e.g. table \ref{gen-errors}.
A star (*) denotes that the quoted forecasted SNR is not ringdown-specific. For ET/CE we have quoted the ringdown-specific ET forecast \cite{Nakano:2021bbw}, in the current absence (to our knowledge) of an analogous forecast for CE. For LISA, we note that the quoted SNR is significantly larger than typical event SNRs in the LISA Mock Data Challenge which go up to $\sim {\cal O}(10^3)$ \cite{Baghi:2022ucj,MockLISADataChallengeTaskForce:2009wir}, while \cite{Zhang:2021kkh} forecast SNRs up to $\sim {\cal O}(10^5)$ for  (sufficiently nearby and massive) events. This also illustrates that there is still significant variance in the forecasted SNRs relevant for the missions considered here.
}
    \label{tab-SNRringdown}
\end{table}

In the previous section we derived parametric expressions for the precision with which the parameters controlling the behaviour of $\aT$ and hence $\cgw$ can be measured for probes with a general ringdown SNR $\rho$. In this section we now summarise and briefly discuss what this concretely implies for a range of current and future missions, spanning the frequency range from $10^{-4}$ Hz to $10^3$ Hz. The main results are collected in table \ref{tab-SNRringdown}.\footnote{It is worth pointing out that the ringdown SNRs quoted implicitly depend on when precisely the transition from merger to ringdown phase is assumed to take place. While discussing this in detail is beyond the scope of this paper, we point the interested reader to \cite{Berti:2007fi,Kamaretsos:2012bs,London:2014cma,LIGOScientific:2016lio,Baibhav:2017jhs,Thrane:2017lqn,Bhagwat:2017tkm} for discussions on this.}

Before discussing forecasted constraints in detail for the respective missions and frequency bands, this is a good point to recall our introductory discussion in section \ref{sec-intro} about how and where the frequency-dependence in $\cgw$ may be localised and how this is tied to the regime (i.e. frequency range) where the underlying theoretical framework is valid.
Rather obviously, any prediction derived from our starting point -- the Horndeski scalar-tensor action \eqref{S} -- is only trustworthy when \eqref{S} is a faithful description of the relevant physics. Since \eqref{S} gives rise to a frequency-independent $\cgw$, we are therefore implicitly assuming that at the very least in the frequency-window spanning the ringdown frequencies in question, $\cgw$ is constant as a function of frequency to high accuracy. A natural scenario to consider would therefore be the one alluded to in the introduction: $\cgw$ effectively becomes a constant $\cgw^{(0)} \neq c$ at low frequencies where \eqref{S} applies and may indeed be intimately linked to dark energy phenomenology on cosmological scales. Now we consider the ringdown following a SMBH merger observable in the LISA band and effectively have $\cgw = \cgw^{(0)}$ there. We can therefore straightforwardly use \eqref{S} to compute this ringdown signal. In this scenario we also assume \eqref{S} stops being an accurate description of the relevant gravitational physics between the LISA and LVK bands and its unknown UV (high energy) completion takes over there, resulting in a transition back to $\cgw = c$ at high frequencies due to the Lorentz invariant nature of the UV completion. The frequency-dependence in $\cgw$, induced by the UV completion is sharply localised in frequency-space between the LISA and LVK bands and so fully consistent with existing bounds on $\cgw$ from the LVK band. Now this scenario -- as explored in detail in the context of forecasting upcoming multi-band constraints in \cite{Harry:2022zey,Baker:2022eiz} -- is only illustrative and the frequency-dependence of $\cgw$ and the regime of validity of \eqref{S} can easily be altered depending on the UV completion and if the connection to dark energy is loosened or severed completely. 
We refer to \cite{deRham:2018red,Baker:2022rhh,Harry:2022zey,Baker:2022eiz} and \cite{Noller:2019chl} for more detailed discussions of those two points, respectively, and note that in this paper this is especially relevant in the context of forecasts for frequency-bands above (i.e. at higher frequencies than) the LISA band. We will come back to this point below.  

What would it take to extrapolate/extend the results from the above sections to cases where $\cgw$ is frequency-dependent in the frequency-window associated with ringdown signals of interest? On the theoretical side, we already pointed out that this would involve supplementing/replacing \eqref{S} with the interactions inducing the frequency-dependence of $\cgw$, which requires knowledge of (or assumptions about) the UV completion of \eqref{S}. The resulting action could then be used to repeat the analysis for this frequency-window. It is worth highlighting that the results of sections \ref{sec:QNMs} and \ref{sec:param-constraints} only know about the Horndeski scalar-tensor action by assuming the corresponding modified form of the Regge-Wheeler equation \eqref{Rweq}. So any UV completion that does not modify this form other than inducing a frequency-dependent $\cgw$ and hence $\aT$ is covered by the analysis in sections \ref{sec:QNMs} and \ref{sec:param-constraints}. We leave an exploration of how UV completions might otherwise  affect the modified Regge-Wheeler equation and how this affects the subsequent analysis for future work. 
On the observational side, a frequency-dependent $\cgw$ would introduce another challenge in the ringdown analysis. Since different parts (i.e. frequencies) of the waveform then travel at different speeds, the received signal at the detector will be stretched/squeezed/scrambled with respect to the signal emitted at the source \cite{Aurrekoetxea:2022ika,Harry:2022zey} -- also see formally related discussions in \cite{Sperhake:2017itk,Rosca-Mead:2020ehn}. Specifically in the ringdown context, this can make identifying the correct frequencies more challenging and this therefore requires a dedicated analysis \cite{Aurrekoetxea:2022ika}.\footnote{We thank Josu Aurrekoetxea and Pedro Ferreira for related discussions.} In practice this means that the strain functions \eqref{strain} accurately describe the signal at emission but will be altered via non-trivial dispersion effects by the time they reach the detector, so this needs to be taken into account to correctly forecast constraints when a frequency-dependent $\cgw$ affects the frequency-window associated with the signal under investigation. We will leave such a dedicated analysis to future work and (as also motivated by the theoretical considerations above) in this section forecast constraints for different frequency bands, assuming an effectively frequency-independent $\cgw$ within the band under investigation (i.e. the LISA forecasts assume a frequency-independent $\cgw$ in the LISA band and so on).

\subsection{LISA band forecasts}

As motivated above and in the introduction, the LISA band is particularly promising in terms of testing for deviations of $\cgw$ from the speed of light, given that a frequency-dependent transition of $\cgw$ just or somewhat below the LVK band is a natural prediction in a range of candidate dark energy models. In terms of the amplitude parameters $A_{Ti}$, we see from table \ref{tab-SNRringdown} 
that one expects the leading order such parameter to be constrained at the $10^{-3}$ level with future LISA/TianQin observations that are forecasted to yield a ringdown SNR of $\sim {\cal O}(10^5)$ \cite{Flanagan:1997sx,Zhang:2021kkh}. Mapping this back to $\aT$ itself \eqref{aTfull}, this implies one will be able to detect deviations down to the $\aT \sim {\cal O}(10^{-4})$ level from LISA band ringdown alone in the context of the models we consider,\footnote{Note that, $\aT$ is generically about one order of magnitude smaller than the dominant $A_{Ti}$ -- see figure \ref{fig-aT}. Also, as should be obvious from \eqref{aTfull}, this mapping is mildly dependent on the $i$ coefficient.} i.e.
\begin{align}
    \sigma_{\aT}^{\rm LISA/TianQin} \sim 10^{-4}.
\end{align}
Note that present forecasts for far-future missions such as AMIGO predict the same order-of-magnitude ringdown SNR, so this would not qualitatively alter constraints on $\cgw$ in comparison with those expected from LISA/TianQin for a single event. 

It is worth emphasising that the main bounds discussed here are forecasted for a single ringdown observation with the SNR achievable by the relevant detector. It is reasonable to expect that qualitatively improved constraints will be obtained when combining multiple observations.
Indeed, for sufficiently large $N$ (where $N$ is the number of detected events) the measurement precision for QNMs is expected to improve as $N^{-1/2}$ \cite{Yang:2017zxs,Brito:2018rfr}, \black{assuming $N$ identical events}. For the LISA band, expected event rates are somewhat uncertain, but most estimates lie in the ${\cal{O}}(10-100)$ per year range for SMBH mergers -- see e.g. \cite{Berti:2006ew,Sesana:2004gf,Rhook:2005pt,Tanaka:2008bv,Berti:2009kk,eLISA:2013xep,Bonetti:2018tpf,Erickcek:2006xc}. An improvement of up to two orders of magnitude on the above constraints therefore seems achievable after several years of operation, so that one may hope to ultimately reach a precision of close to $\sigma_{\aT}^{\rm LISA/TianQin} \sim 10^{-6}$. \black{We again emphasise that the $N^{-1/2}$ scaling discussed here assumes an idealised case with $N$ identical events with the large SNRs considered here and so the above should be taken as an optimistic bound (e.g. many events to be detected will be at higher redshifts and have correspondingly reduced SNRs), also depending on the precise SMBH merger and detection rates as discussed above.}

\subsection{LVK band forecasts} 

Having summarised results for the LISA band above, let us consider the LVK band. The situation here is qualitatively different, given that there already are tight constraints on $\cgw$ specific to this frequency band. From measuring the coincidence of the GW170817 signal in GW and optical counterpart observations, one finds that $\aT \lesssim 10^{-15}$ \cite{TheLIGOScientific:2017qsa,2041-8205-848-2-L14,2041-8205-848-2-L15,LIGOScientific:2017zic,LIGOScientific:2017ync}. When even a very mild frequency-dependence of $\cgw$ is present in the LVK band, this bound can be strengthened to $\aT \lesssim 10^{-17}$ \cite{Harry:2022zey}. 
Contrast this with bounds from ringdown observations alone, where the $A_{Ti}$ can be constrained at the $\sim {\cal O}(10)$ level with LVK observations with an improvement by approximately an additional order of magnitude to be expected from the future Einstein Telescope(ET)/Cosmic Explorer(CE) missions -- cf. table \ref{tab-SNRringdown} and see \cite{Maggiore:2019uih,Nakano:2021bbw} and \cite{Evans:2021gyd,Hall:2022dik} for ET and CE, respectively. Again mapping this to constraints on $\aT$ itself, we therefore ultimately expect
\begin{align}
    \sigma_{\aT}^{\rm ET/CE} \sim 10^{-1}.
\label{sigma_aT_ETCE}
\end{align}
once ET/CE are collecting data in the future. 

The bound \eqref{sigma_aT_ETCE} given above is again for a single event with the SNR achievable by ET/CE. Taking into account the $N^{-1/2}$ improvement of the measurement precision for $N$ detected events discussed above, we can again extrapolate how this precision might be improved over time. For ET ${\cal{O}}(10^4-10^5)$ events with a ringdown SNR of ${\cal O}(10)$ are expected per year \cite{Berti:2016lat}. One may therefore reasonably expect that constraints can eventually be improved by about two orders of magnitude to $\sigma_{\aT}^{\rm ET/CE} \sim 10^{-3}$. In the LVK context it is also interesting to point out that existing (non-ringdown-specific) constraints on $\cgw$ from GW waveforms in the LVK band have already seen similar improvements by stacking events. More specifically, when comparing I) constraints on $\cgw$ from GW170817 data alone (i.e. without using an optical counterpart) \cite{LIGOScientific:2018dkp} with II) constraints obtained using a LVK catalog of 43 confident binary black hole mergers (used to obtain bounds on the graviton mass in~\cite{LIGOScientific:2021sio}, but straightforwardly re-interpretable to place bounds on $\cgw$), this improves these bounds on $\cgw$ by around two orders of magnitude.\footnote{This improvement, while still partially driven by the larger number $N$ of observations included, is also partially due to other events having higher individual SNRs than GW170817.} 


At first sight \eqref{sigma_aT_ETCE} as well as the improved $\sigma_{\aT}^{\rm ET/CE} \sim 10^{-3}$ bound reachable by stacking events are rather weak, albeit complementary, constraints on $\cgw$ when compared with the existing GW propagation bounds from GW170817 discussed above.  
Also note that, for the purposes of this subsection and as discussed in detail above, we are assuming that \eqref{S} is a valid description of the underlying physics in (at least part of) the LVK band. 
As discussed, in dark energy-related theories within \eqref{S} where $\cgw$ receives order one corrections on cosmological scales one would not expect this to be the case. \black{One can remedy this (i.e. `return' the LVK band to within the regime of validity of \eqref{S}) in two different ways. First, by severing the connection to cosmology/dark energy and looking at the constraints derived here in their own right. Or second, by suppressing the cosmological $\aT$ from the beginning while not precluding a more sizeable $\aT$ around black hole space-times.} We will briefly recap a specific scenario related to the second case below.
However, a more general related point is the following: The fact that the constraints derived in this paper are computed for a different background solution than cosmological background GW-propagation constraints derived e.g. from GW170817 means that they nevertheless contain some interesting new information on the scalar hair profile and the underlying interactions encoded in $G_{T}$ along the lines discussed in section \ref{subsec-HairProfile} -- we show this in figure \ref{errorSNRphi}. More specifically, from \eqref{sigmaGT} and in the event of a scalar hair amplitude $\varphi_c \sim {\cal O}(10^8)$, the constraint on the underlying interactions will be as strong as constraints on the same interactions from GW170817 and even stronger for a larger amplitude $\varphi_c$. 
Reversing the argument, if future observations were to identify a small but non-zero cosmological $\aT$, this would allow placing a bound on the scalar hair amplitude from the ringdown constraints investigated here. For concreteness, consider the following setup: The higher derivative scalar interactions in the $G_{3,4,5}$ terms in \eqref{S} come with an implicit mass scale $\Lambda$. In cosmology this scale is typically chosen to be $\Lambda = \Lambda_3 \equiv (\MPl H_0^2)^{1/3}$, where this choice ensures those interactions give ${\cal O}(1)$ contributions to cosmology. However, if a different $\Lambda$ is chosen, the cosmological $\aT$ (and hence $G_T$) scales as $\alpha_T \sim (\Lambda_3/\Lambda)^6$. So raising the interaction scale $\Lambda$ by just three orders of magnitude suppresses the cosmological $\aT$ down to a level of ${\cal O}(10^{-18})$, comfortably consistent with bounds from GW170817. This setup also allows the full LVK band to be within the regime of validity of the physics described by \eqref{S} -- see \cite{Noller:2019chl} for further details on this scenario. Now suppose that a future constraint indeed establishes $G_T \sim {\cal O}(10^{-18})$, while future ringdown constraints from ET/CE along the lines investigated here do not yield evidence for a non-zero $\aT$. From \eqref{ATi_def} this would allow us to derive a bound on the scalar hair $\varphi_c \lesssim {\cal O}(10^9)$ for frequencies in the LVK band. Note that other complementary bounds on $\varphi_c$ may be obtainable e.g. from considering even perturbations or going beyond linear theory.

\subsection{Intermediate band forecasts} 

With LISA and LVK forecasts discussed above, the intermediate frequency band stands out as a third region of interest. Here the upcoming AEDGE \cite{AEDGE:2019nxb} and DECIGO \cite{Kawamura:2006up,Kawamura:2020pcg} experiments will detect and investigate GWs in the future. In the introduction we motivated probing $\cgw$ in the LISA band by pointing out that a frequency-dependent transition from a nearly constant $\cgw = c$ at LVK frequencies to a different low-frequency $\cgw$ naturally occurs just or somewhat below the LVK band in large classes of dark energy theories. This motivation of course equally applies to the frequencies probed by AEDGE/DECIGO. Candidate transitions in this intermediate band may `leak out' into the LISA and/or LVK bands, in which case the considerations outlined above for those bands already promise tight constraints. However, another interesting class of transitions are those investigated by \cite{Harry:2022zey,Baker:2022eiz}, where the transition is effectively completely contained  within the intermediate frequency band and no detectable frequency-dependence leaks out into the LISA and/or LVK bands. In such a case multiband observations using systems such as GW150914 that are first observable in the LISA band and eventually enter the LVK band can be used to obtain an integrated constraint on any features residing at intermediate frequencies and indeed will be able to constrain $\aT$ down to a level of ${\cal O}(10^{-15})$ \cite{Harry:2022zey,Baker:2022eiz}. 
In addition, once AEDGE/DECIGO observations are available, direct constraints on $\cgw$ from this band will be obtainable in analogy to the LVK/LISA analyses discussed above. 
Whenever there is significant frequency-dependence for $\cgw$ in band, a complementary ringdown-specific analysis faces similar theoretical challenges as discussed for the LVK band above, as well as the observational modelling challenges mentioned earlier in this section. 
So, as before, the bounds forecasted in this subsection will be for the case where \eqref{S} applies within (at least part of) the AEDGE/DECIGO band and hence $\cgw$ is frequency-independent in this band to high accuracy.
With these assumptions and from ringdown alone, we find that AEDGE/DECIGO will be able to constrain the $A_{Ti}$ at the $\sim {\cal O}(10^{-1})$ level, c.f. table \ref{tab-SNRringdown}. Mapping this to constraints on $\aT$ itself, as before, this implies
\begin{align}
    \sigma_{\aT}^{\rm AEDGE/DECIGO} \sim 10^{-2}.
\label{sigma_aT_AEDE}
\end{align}
While weaker in magnitude than the integrated multiband constraints discussed above, these bounds are complementary in the same sense as discussed in the LVK section above. Note that one may again expect this bound to be improved significantly when stacking multiple observed events: Several dozen intermediate mass black hole (IMBH) mergers with an SNR ${\cal O}(10^3)$ should be observable with AEDGE per year \cite{AEDGE:2019nxb}, so optimistically an improvement up to $\sigma_{\aT}^{\rm AEDGE/DECIGO} \sim 10^{-4}$ appears feasible eventually.

\section{Conclusions} \label{sec:conclusions}
In this paper we have investigated how the speed of gravitational waves $\cgw$ can be probed using black hole ringdown observations. Focusing on scalar-tensor theories of the Horndeski type and on odd parity quasinormal modes (QNMs), our key findings are as follows:
\begin{itemize}
    \item In the context of non-rotating black holes where the metric background solution is given by Schwarzschild, we find that deviations of $\cgw$ from the speed of light only affect the QNMs in the presence of a non-trivial scalar hair profile $\phi=\phi(r)$ in agreement with the results of \cite{Tattersall:QNMHair}. Any deviations from $\cgw = c$ are then proportional to the square of the amplitude of the scalar hair. 
    \item For a single event, ringdown observations from LISA and TianQin will be able to constrain $\cgw$ at the ${\cal O}(10^{-4})$ level.  For AEDGE/DECIGO the equivalent precision will be ${\cal O}(10^{-2})$. When stacking observations over several years, both constraints may be improved by up to two orders of magnitude, depending on precise event rates. While those constraints are weaker than existing constraints on $\cgw$ e.g. from GW170817, the  importantly probe different frequency ranges. This is particularly relevant in the context of testing $\cgw$, given large classes of dark energy models naturally give rise to a frequency-dependent transition in $\cgw$ below the LVK band. 
    \item With ringdown constraints we are testing the effect of deviations from $\cgw=c$ on a different background solution than that relevant for GW propagation constraints on $\cgw$ (black hole vs. cosmological space-times). \black{The precise dependence of $\cgw$ on interactions in the underlying theory is different for these two backgrounds. We have highlighted examples where, in the presence of a sufficiently large scalar hair profile, ringdown observations can provide novel constraints on those interactions.} Likewise, given complementary information on those underlying interactions, we have shown how ringdown observations can constrain the nature of scalar hair. We stress that therefore even LVK band ringdown measurements, where we find that ${\cal O}(10^{-1})$ level will be obtainable from the Einstein Telecope/Cosmic Explorer for a single event, can yield valuable information complementary to existing constraints on $\cgw$.  
\end{itemize}
Overall we have therefore derived forecasts for the precision with which ringdown observations will be able to constrain the speed of gravitational waves $\cgw$ for various detectors throughout the ${\cal O}(10^{-4})-{\cal O}(10^3)$ Hz frequency range. 
Our study has been idealised in the sense that we have assumed I) the `usual' binary black hole merger parameters (masses, amplitudes, phases) to be known and focused on the effect of novel parameters associated with $\cgw \neq c$, II) focused on a specific background solution for the black hole geometry and scalar hair profile, and III) by working with Horndeski scalar tensor theories, we have implicitly assumed that $\cgw$ is approximately constant in specific frequency windows/bands when forecasting constraints for those respective bands.
A more comprehensive analysis, extending the present work, investigating degeneracies and constraints in higher-dimensional parameter spaces as well as a wider range of hairy black hole solutions and underlying theoretical setups, will therefore be an interesting next step.
As has been mentioned, another promising route to make our setup more physically realistic is to extend the quasinormal mode calculations to rotating black hole solutions \cite{Tattersall:KerrdSBH,Pani:2012bp,doi:10.1142/S0217751X13400186,Pani:2013wsa,Brito:2013wya}. Another step in this direction would be to include in the analysis surrounding matter fields that dynamically interact with the black hole in a way that also affects the emitted quasinormal modes -- see e.g.  \cite{Bamber:2021knr,Bamber:2022pbs,Leung:1997was,PhysRevD.75.044016,Medved:2003pr,Barausse:2014tra,https://doi.org/10.1002/asna.201913573,Matyjasek:2020bzc}.
It is also worth emphasising that there are several complementary probes of $\cgw$ in addition to the gravitational wave probes discussed throughout this paper and corresponding to energy/frequency scales outside of the range considered here. These include constraints from cosmological large scale structure, currently at the ${\cal O}(1)$ level -- see e.g. \cite{Bellini:2015xja,Noller:2018wyv} and references therein -- which are expected to improve to ${\cal O}(10^{-1})$ in the near future \cite{Alonso:2016suf}. 
While we have not mandated a specific sign for any potential deviation of $\cgw$ away from $c$, theoretical bounds from requiring causality, locality and unitarity at high energies can further yield information on these deviations at the (comparatively) low energies probed by gravitational waves and cosmology, noticeably mandating $\cgw \geq c$ for large classes of models \cite{deRham:2019ctd,Melville:2019wyy}.
We close by re-emphasising that we have mostly focused on investigating how well $\cgw$ can be tested by ringdown observations, for a single detected odd-parity quasinormal mode. As more sources and modes are detected in the future and the theoretical machinery to analyse them is further developed, we fully expect further tightened constraints to become obtainable. 

\section*{Acknowledgments}
\vspace{-0.1in}
We thank Josu Aurrekoetxea, David Bacon, Emanuele Berti, Pedro Ferreira, Ian Harry, Kazuya Koyama, Andrew Lundgren, Sigurd Naess, Thomas Sotiriou and Oliver Tattersall for useful discussions. SSL is supported by an STFC studentship. JN is supported by an STFC Ernest Rutherford Fellowship (ST/S004572/1). In deriving the results of this paper, we have used xAct~\cite{xAct} as well as the xAct notebook repository by Reginald Bernardo \cite{reggierepo,Bernardo:2021vsj,Bernardo:2020ehy}.

\appendix

\section{Parametrised hair}\label{app-full-param}
In this appendix we quote the results of \cite{Tattersall:QNMHair}, which consider the full parametrised hair ansatz (\ref{hair_parametrisation}). The modified Regge-Wheeler equation becomes

\begin{widetext}
\begin{align}
\left[\frac{d^2}{dr_\ast^2}+\omega^2\left(1+\epsilon^2\alpha_{T}(r)\right)-A(r)\left(\frac{\ell(\ell+1)}{r^2}-\frac{6M}{r^3}+\epsilon\delta V_1+\epsilon^2\delta V_2 \right)\right]Q=0,
\label{RWhairy}
\end{align}
\end{widetext}
where the potential perturbations are given by
\begin{widetext}
\begin{subequations}\label{deltaVs}
\begin{align}
\delta V_1=&\;\frac{1}{2r^2}\left[4\delta A_1-2r\delta A_1^\prime-2(\ell+2)(\ell-1)\delta C_1+2(r-3M)\delta C_1^\prime-r(r-2M)\delta C_1^{\prime\prime}-\frac{G_{4\phi}}{G_4}\left(r\left(r-2M\right)\delta\phi_1^{\prime\prime}-2(r-3M)\delta\phi_1^\prime\right)\right]\label{deltaV1}\\
\delta V_2=&\;\frac{1}{4r^2}\left[8\delta A_2-4r\delta A_2^\prime+4(\ell+2)(\ell-1)\left(\delta C_1^2-\delta C_2\right)+3r(r-2M)\delta C_1^{\prime2}+4(r-3M)\delta C_2^\prime-2r(r-2M)\delta C_2^{\prime\prime}+4r\delta A_1\delta C_1^\prime\right.\nonumber\\
&\left.-2r^2\left(\delta A_1^\prime \delta C_1^\prime+\delta A_1\delta C_1^{\prime\prime}\right)-4(r-3M)\delta C_1\delta C_1^{\prime\prime}+2r(r-2M)\delta C_1 \delta C_1^{\prime\prime}\right]\nonumber\\
&-\frac{1}{2r^2}\frac{G_{4\phi}}{G_4}\left[-2(r-3M)\delta\phi_2^\prime + r \left( r\delta A_1^\prime\delta\phi_1^\prime -\delta A_1\left(2\delta \phi_1^\prime - r \delta\phi_1^{\prime\prime}\right)+(r-2M)\left(\delta\phi_2^{\prime\prime}-\delta C_1^\prime \delta\phi_1^\prime\right)\right)\right]\nonumber\\
& + \frac{1}{4r^2}\left(\frac{G_{4\phi}}{G_4}\right)^2\left[3r(r-2M)\delta\phi_1^{\prime 2}+2\delta\phi_1\left(r(r-2M)\delta\phi_1^{\prime\prime}-2(r-3M)\delta\phi_1^\prime\right)\right]\nonumber\\
& - \frac{1}{2r^2}\frac{G_{4\phi\phi}}{G_4}\left[r(r-2M)\delta\phi_1^{\prime 2}+\delta\phi_1\left(r(r-2M)\delta\phi_1^{\prime\prime}-2(r-3M)\delta\phi_1^\prime\right)\right]\nonumber\\
& - \frac{\alpha_T(r)}{2r^3}\left[-5M+Mr(r-2M)^{-1}-2r(\ell+2)(\ell-1)+r^2(r-2M)\left(\frac{\delta\phi_1^{\prime\prime}}{\delta\phi_1^\prime}\right)^2 + r\left(r(r-2M)\frac{\delta\phi_1^{\prime\prime\prime}}{\delta\phi_1^\prime}-2(r-5M)\frac{\delta\phi_1^{\prime\prime}}{\delta\phi_1^\prime}\right)\right].\label{deltaV2}
\end{align}
\end{subequations}
\end{widetext}
There are several observations we can make from these expressions. Let us first point out that the modified Regge-Wheeler equation in our main text \eqref{Rweq} with \eqref{dV} can be recovered by setting $\delta A_{1,2}=\delta C_{1,2}=G_{4\phi}=G_{4\phi\phi}=0$. However, we see that $\delta A_1, \delta C_1$ would contribute at lower order (as well as $\delta\phi_1$, if $G_{4\phi} \neq 0$). 
If present, such terms can therefore significantly contribute to the ringdown signal. Indeed, if the fiducial scalar hair is highly suppressed, i.e. when $\epsilon$ is very small, such lower-order-in-$\epsilon$ contributions would be expected to dominate over any $\aT$-induced contributions. 
The motivation behind our simple setup then is not to fully explore all parametric effects and degeneracies in a comprehensive parameter space, but
rather to isolate and investigate observable signatures of $\aT$. Having said this, it has been shown that for known scalar-tensor theories that have hairy black holes, namely scalar-Gauss-Bonnet theory, the metric at leading perturbative order remains Schwarzschild (i.e. $\delta A_{1}=\delta C_{1}=0$) \cite{Sotiriou:2014pfa}, suggesting that several aspects of our simple setup are concretely realised in relevant theories.

In addition, one could consider cases where $G_{4\phi}\neq0\neq G_{4\phi\phi}$. It is interesting to point out that interactions contributing to $G_{4\phi}$ in our background can be removed with a conformal transformation of the metric. This is because we still have $\widehat X=0$,\footnote{Recall that $\widehat X$ refers to the kinetic term $X$ evaluated on the `reference background' of $\widehat \phi$ (i.e. $\delta\phi=0$).} meaning that all terms in $G_{4\phi}$ are $X$-independent or, in other words, those interactions are of the form $f(\phi)R$ in Jordan frame, which is well known to be convertible to the usual Einstein-Hilbert term by a conformal transformation. In the resulting Einstein frame representaton, one then naturally finds $G_{4\phi}=0=G_{4\phi\phi}$. This would indeed make our assumptions more general and, because the Horndeski group is closed under conformal transformations and we are not including matter fields, our calculations would follow exactly in the same way. The price to pay for working with the metric in the Einstein frame is that observations in GW detectors are coupled to matter and therefore measure the metric in Jordan frame. Hence, forecasts should ultimately be made for gravitational waves as observed in the detector/Jordan frame. We therefore abstain from removing any interactions via a conformal transformation here and explicitly highlight setting $G_{4\phi}=0=G_{4\phi\phi}$ as an additional simplifying assumption. Note that this assumption is trivially satisfied in theories where the scalar $\phi$ is endowed with a shift symmetry $\phi \to \phi + c$.

To conclude this appendix, let us briefly consider the case where one or many of $G_{4\phi}, G_{4\phi\phi},\delta A_{2},\delta C_{2}$ are non-vanishing. In this case an analogous analysis to the one performed in this paper can still be carried out, where the functional form of any non-vanishing such functions would need to be specified as above. The additional parameters introduced in this way mean a higher parameter space would then have to be constrained, presumably degrading constraints on individual parameters. Breaking degeneracies in such a higher-dimensional parameter space would likely require the measurements of multiple QNMs. We leave such a more comprehensive exploration to future work.

\section{Coefficients in quadratic action}
\label{app-acoefficients}
In this appendix we quote the coefficients in the quadratic action for a general $A$, $B$ and $C$, as derived in \cite{Kobayashi:2012kh,Ganguly:2017ort}.
The quadratic action is given by
\begin{align}
    S^{(2)}=\int dtdr\left[\bar a_1h_0^2+\bar a_2h_1^2+\bar a_3(\dot{h}_1^2+h_0'^2-2\dot{h}_1h_0'+2\frac{C'}{C}\dot{h}_1h_0)
    \right]
\end{align}
with the coefficients being
\begin{align}
    &\bar a_1=\frac{\ell(\ell+1)}{4C}\left[\left(C'\sqrt{\frac{B}{A}}\mathcal{H}\right)'+\frac{(\ell-1)(\ell+2)}{\sqrt{AB}}\mathcal{F}+\frac{2C}{\sqrt{AB}}\epsilon_A\right],\nn\\
    &\bar a_2=-\frac{\ell(\ell+1)}{2}\sqrt{AB}\left[\frac{(\ell-1)(\ell+2)}{2C}\mathcal{G}+\epsilon_B\right],\nn\\
    &\bar a_3=\frac{\ell(\ell+1)}{4}\sqrt{\frac{B}{A}}\mathcal{H},
\end{align}
where again the bar denotes that the $\bar a_i$ are evaluated on the background and, to avoid clutter, bars are implied everywhere on the right hand side. $\epsilon_{A,B}$ are contributions that vanish on-shell\footnote{Expressions for $\epsilon_{A,B}$, which are obtained by varying the action with respect to $A$ and $B$, are fully provided in \cite{Ganguly:2017ort}.} and
\begin{align}
    \mathcal{F}&=2\left(G_4+\frac{1}{2}B\phi'X'G_{5X}-XG_{5\phi}\right),\nn\\
    \mathcal{G}&=2\left[G_4-2XG_{4X}+X\left(\frac{A'}{2A}B\phi'G_{5X}+G_{5\phi}\right)\right],\nn\\
    \mathcal{H}&=2\left[G_4-2XG_{4X}+X\left(\frac{C'}{2C}B\phi'G_{5X}+G_{5\phi}\right)\right].
\end{align}
As shown in \cite{Kobayashi:2012kh,Ganguly:2017ort}, this action can be rewritten to make the presence of only one degree of freedom explicit
\begin{align}
    S^{(2)}=\frac{\ell(\ell+1)}{4(\ell-1)(\ell+2)}\int dtdr_*\left[\frac{\mathcal{F}}{\mathcal{G}}\dot{Q}^2-\left(\frac{dQ}{dr^*}\right)^2-V(r)Q^2\right], \nn \\ 
\end{align}
where the potential is given by
\begin{align}
    V &= \ell(\ell+1)\frac{A}{C}\frac{\mathcal{F}}{\mathcal{H}}-\frac{C'^2}{4C''}\left(\frac{ABC'^2}{C^3}\right)' \nn \\ &\;\;\; \;-\frac{C^2\mathcal{F}^2}{4\mathcal{F}'}\left(\frac{AB\mathcal{F}'^2}{C^2\mathcal{F}^3}\right)'-\frac{2A\mathcal{F}}{C\mathcal{H}}.
\end{align}
Our quadratic action \eqref{S2v2} and its coefficients as well as the potential \eqref{V} can be recovered from the expressions above by specifying our background: $A=B$, $C=r^2$.

\section{General scalar profile}\label{app-generaln}
Here we repeat the derivation of quasinormal corrections but for a more general scalar profile, given by
\begin{equation}
    \delta\phi=\varphi_c\left(\frac{2M}{r}\right)^n,
    \label{gen-scalar-prof}
\end{equation}
which means $\aT$ is now given by
\begin{align}
    \alpha_T= -\sum_{i=0}^\infty A_{Ti}\cdot n^2\left(1-\frac{2M}{r}\right)\left(\frac{2M}{r}\right)^{2n+i+2},
    \label{aTin}
\end{align}

\begin{figure}[t!]
    \centering
    \includegraphics[width = 0.48\textwidth]{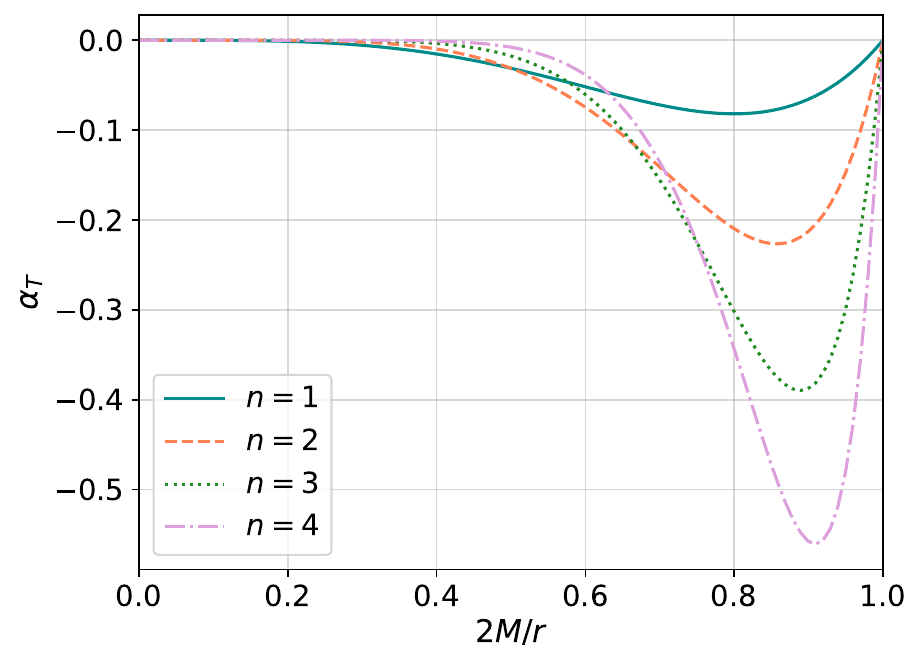}\\[0.5cm]
    \caption{Here we show $\aT$ \eqref{aTin} for different choices of $n$ in \eqref{gen-scalar-prof} as a function of $2M/r$. $i=0$ has been set such that $A_{T0}$ is the only non-zero parameter (and fixed to a fiducial value of $1$). We see that $\aT(r) = 0$ at spatial infinity and at the horizon, but again observe non-trivial behaviour in the intermediate region. The size of $\aT$ (partially controlled by $A_{T0}$) is considerably enhanced by increasing $n$. This is mainly due to the factor $n^2$ accompanying all $\delta\omega$ (as can be seen from \eqref{domen} and \eqref{E-basis}). 
    }
    \label{alT-n}
\end{figure}
\begin{figure*}
    \centering
    \includegraphics[width = 0.8\textwidth]{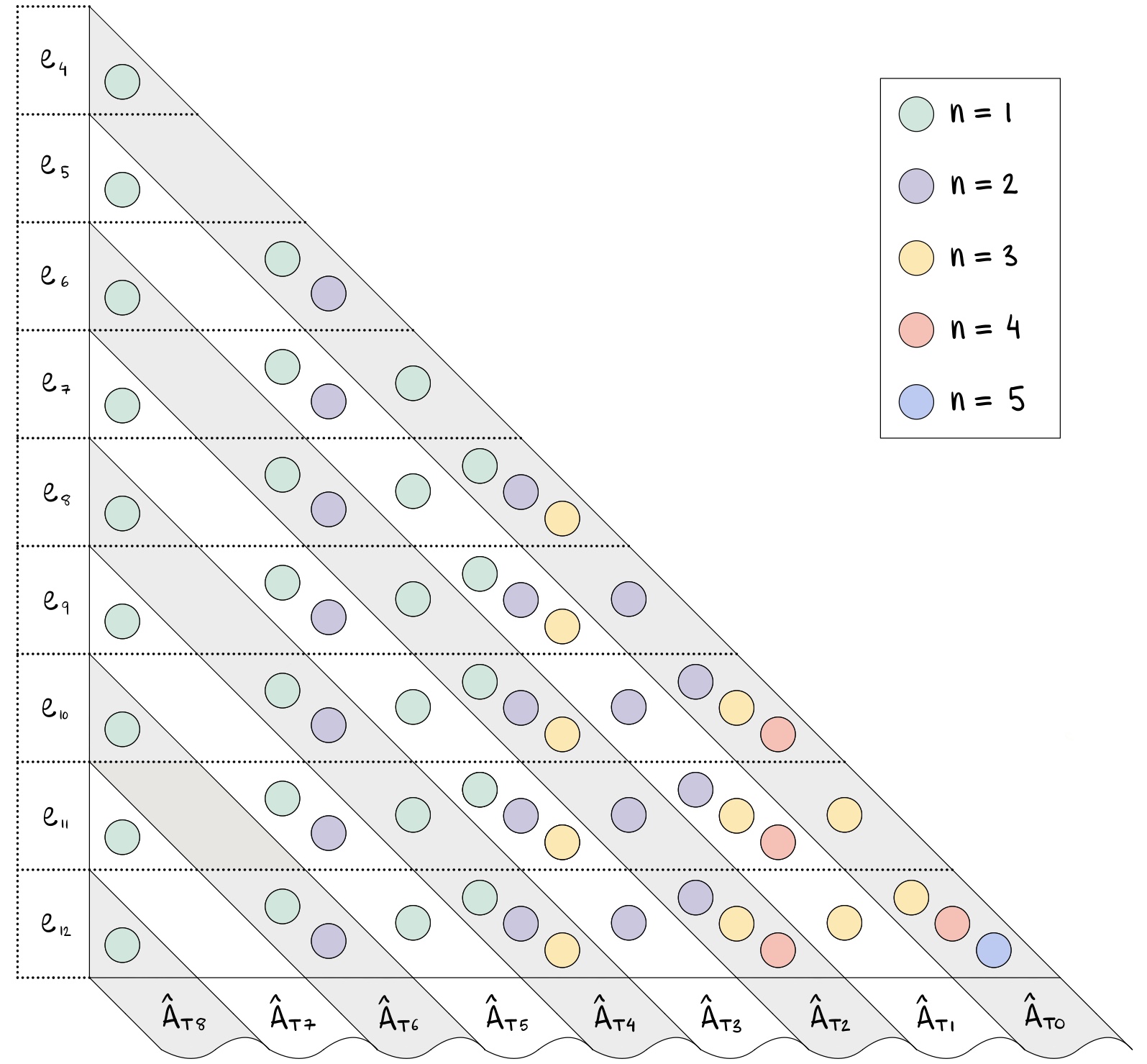}\\[0.5cm]
    \caption{In this diagram we display the contributions to $\alpha_T$ \eqref{aTin} coming from different $n$ values. These are ordered such that all contributions to a specific $A_{Ti}$ parameter appear in one diagonal, which in principle is extendible \textit{ad infinitum}. Contributions from different $n$ also follow a pattern, which can be appreciated here by focusing on the colour of the balls. The vertical axis $e_j$ corresponds to the paramtrized basis from \cite{ParamRingdown}, which tells us the power of $(1/r)^j$ at which each contribution appears in the potential.}
    \label{triangle}
\end{figure*}

where the new $n$-dependence is plotted in figure \ref{alT-n}. Substituting this back into $\tilde{\delta V}$ (\ref{dVtilde}) we get
\begin{equation}
\begin{split}
   \tilde{\delta V}=\left(\frac{n}{2M}\right)^2&\sum_{i=0}^\infty A_{Ti}\Bigg[\left(\frac{2M}{r}\right)^{2n+i+2}(2M\omega_0)^2\\
   +&\left(\frac{2M}{r}\right)^{2n+i+4}\left(-\ell(\ell+1)+\frac{9}{2}+\frac{7}{2}n+n^2\right)\\
   +&\left(\frac{2M}{r}\right)^{2n+i+5}\left(\ell(\ell+1)-\left(\frac{19}{2}+\frac{17}{2}n+2n^2\right)\right)\\
   +&\left(\frac{2M}{r}\right)^{2n+i+6}\left(\frac{21}{4}+5n+n^2\right)\Bigg].
\end{split}
\end{equation}
Note that setting $n=1$ recovers the expression \eqref{dVn1}. Again, this can be written as
\begin{equation}
    \tilde{\delta V}=\frac{1}{(2M)^2}\sum_{j=0}^{\infty}a_j\left(\frac{2M}{r}\right)^j
\end{equation}
with the only non-zero a-parameters for a given $i$ contributing as
\begin{align}
    &a_{2n+i+2}\ \ \backin\ \ A_{Ti}\cdot n^2(2M\omega_0)^2\nn\\
    &a_{2n+i+4}\ \ \backin\ \ A_{Ti}\cdot n^2\left(-\ell(\ell+1)+\frac{9}{2}+\frac{7}{2}n+n^2\right)\nn\\
    &a_{2n+i+5}\ \ \backin\ \ A_{Ti}\cdot n^2\left(\ell(\ell+1)-\left(\frac{19}{2}+\frac{17}{2}n+2n^2\right)\right)\nn\\
    &a_{2n+i+6}\ \ \backin\ \ A_{Ti}\cdot n^2\left(\frac{21}{4}+5n+n^2\right).
    \label{generalnacoefs}
\end{align}
From these we find the following quasinormal frequency corrections
\begin{equation}
    \delta\omega=\sum_{j=0}^\infty a_j e_j=\sum_{i=0}^\infty A_{Ti} E_i^n,
    \label{domen}
\end{equation}
where we have defined the following new basis for convenience
\begin{align}
    E_i^n=n^2\Bigg[&(2M\omega_0)^2e_{2n+i+2}\nn \\
    &+\left(-\ell(\ell+1)+\frac{9}{2}+\frac{7}{2}n+n^2\right)e_{2n+i+4}\nn \\
    &+\left(\ell(\ell+1)-\left(\frac{19}{2}+\frac{17}{2}n+2n^2\right)\right)e_{2n+i+5}\nn \\
    &+\left(\frac{21}{4}+5n+n^2\right)e_{2n+i+6}\Bigg].
    \label{E-basis}
\end{align}
The basis for $n=1$ \eqref{domega_v2} and $n=2$ \eqref{En2} used in the main text can be straightforwardly recovered from this.
Note that here, because we have chosen to remain agnostic about $n$, we have ended up with two indices, i.e. $n$ and $i$ that need to be chosen in order to obtain numerical results. This is shown explicitly in the super and subscripts of the newly defined basis $E_i^n$. In diagram \ref{triangle} we display the corrections coming from different choices of $(i,n)$, and make some observations about their structure.

The step from these analytically calculated corrections to the quasinormal modes to the errors on different $A_{Ti}$ parameters (and hence corrections on $\alpha_T$) is straightforwardly repeated in the same fashion as shown in section \ref{sec:param-constraints}. We show in table \ref{gen-errors} the results for a few more illustrative cases, but stress that such an analysis can easily be repeated for any combination and superposition of $(i,n)$ by adapting the companion notebook provided in \cite{ringdown-calculations}. The general trend we find that can already be appreciated in table \ref{gen-errors} is that errors decrease for increasing $i$ and $n$ (at least for the single-parameter cases).

 \begin{table}[!h]
 \setlength{\tabcolsep}{10pt}
     \centering
     \begin{tabular}{llll}\toprule
                        & $i=0$ & $i=1$ & $i=\{0,1\}$\\\midrule
          $n=1$         & 181   & 131   & (68,\ 550) \\
          $n=2$         & 132   & 72    & (27,\ 250) \\ 
          $n=\{1,2\}$   & 153   & 93    & (40,\ 344) \\\bottomrule
     \end{tabular}
     \caption{Values for $\sigma_{A_{Ti}}\rho$ for a scalar profile $\delta\phi=\varphi_c\left(\frac{2M}{r}\right)^n$. The last column ($i:0,1$) displays the errors for the two parameters $(A_{TA},A_{TB})$, obtained from $(A_{T0},A_{T1})$ via the same diagonalization procedure as shown in section \ref{multATi}.}
     \label{gen-errors}
 \end{table}

\bibliographystyle{utphys}
\bibliography{Horn_QNMs}
\end{document}